\documentclass[12pt]{article}
\usepackage[a4paper, top=16mm, text={170mm, 248mm}, includehead, includefoot, hmarginratio=1:1, heightrounded]{geometry}
\usepackage{amsmath,amssymb,mathrsfs,amsthm,tikz,shuffle,paralist}
\usepackage{dsfont}
\usepackage{color}
\definecolor{darkred}{RGB}{173,34,48}

\usetikzlibrary{snakes}
\usetikzlibrary{calc}
\usetikzlibrary{decorations}

\usepackage[all]{xypic}
\usepackage{jheppub}
\usepackage{hyperref}
\usepackage{subfigure}
\usepackage{caption}

\newcommand{\dif}{\mathrm{d}} 

\DeclareMathOperator{\Li}{Li} 
\newcommand{\llangle}{\langle\!\langle}
\newcommand{\rrangle}{\rangle\!\rangle}

\title{Notes on cluster algebras and some all-loop Feynman integrals}

\date{\today}
\author[a,b,c,d]{Song He}
\author[a,d]{Zhenjie Li} 
 \author[a,d]{Qinglin Yang}%
\affiliation[a]{CAS Key Laboratory of Theoretical Physics, Institute of Theoretical Physics, Chinese Academy of Sciences, Beijing 100190, China}
\affiliation[b]{
School of Fundamental Physics and Mathematical Sciences, Hangzhou Institute for Advanced Study, UCAS, Hangzhou 310024, China}
\affiliation[c]{ICTP-AP
International Centre for Theoretical Physics Asia-Pacific, Beijing/Hangzhou, China}
\affiliation[d]{School of Physical Sciences, University of Chinese Academy of Sciences, No.19A Yuquan Road, Beijing 100049, China}
\emailAdd{songhe@itp.ac.cn}
\emailAdd{lizhenjie@itp.ac.cn}
\emailAdd{yangqinglin@itp.ac.cn}

\abstract{We study cluster algebras for some all-loop Feynman integrals, including box-ladder, penta-box-ladder, and double-penta-ladder integrals. In addition to the well-known box ladder whose symbol alphabet is $D_2\simeq A_1^2$, we show that penta-box ladder has an alphabet of $D_3\simeq A_3$ and provide strong evidence that the alphabet of seven-point double-penta ladders can be identified with a $D_4$ cluster algebra. We relate the symbol letters to the ${\bf u}$ variables of cluster configuration space, which provide a gauge-invariant description of the cluster algebra, and we find various sub-algebras associated with limits of the integrals. We comment on constraints similar to extended-Steinmann relations or cluster adjacency conditions on cluster function spaces. Our study of the symbol and alphabet is based on the recently proposed Wilson-loop ${\rm d}\log$ representation, which allows us to predict higher-loop alphabet recursively; by applying it to certain eight-point and nine-point double-penta ladders, we also find $D_5$ and $D_6$ cluster functions respectively. }
\begin{document}

\maketitle

\section{Introduction and Review}

Recent years have witnessed enormous progress in unravelling rich mathematical structures of scattering amplitudes in QFT, especially for ${\cal N}=4$ super-Yang-Mills (SYM) in the planar limit. The all-loop integrand has been determined using on-shell data~\cite{ArkaniHamed:2010kv} and reformulated geometrically using positive Grassmannian~\cite{Arkani-Hamed:2016byb} and amplituhedron~\cite{Arkani-Hamed:2013jha}. The (integrated) amplitudes have been determined to impressively high loop orders, at least for $n=6$ and $n=7$ ({\it c.f.}~\cite{Dixon:2011pw,Dixon:2014xca,Dixon:2014iba,Drummond:2014ffa,Dixon:2015iva,Caron-Huot:2016owq,Dixon:2016nkn,Drummond:2018caf, Caron-Huot:2019vjl, Caron-Huot:2019bsq, Dixon:2020cnr} and a review~\cite{Caron-Huot:2020bkp}.) The starting point for such bootstrap is the observation~\cite{Golden:2013xva} that the {\it symbol alphabet} of $n=6$ and $n=7$ amplitudes are dictated by $A_3$ and $E_6$ cluster algebras naturally associated with $G(4,6)/T$ and $G(4,7)/T$ respectively. Starting $n=8$, the cluster algebras become infinite and the (finite) symbol alphabet involves algebraic letters which go beyond usual cluster coordinates. Recently, the two-loop NMHV amplitudes have been computed for $n=8$~\cite{Zhang:2019vnm} and higher \cite{He:2020vob} using the method of ${\bar Q}$ equations~\cite{CaronHuot:2011kk}, and the alphabet has been explained using  tropical positive Grassmannian~\cite{Drummond:2019cxm, Henke:2019hve, Arkani-Hamed:2019rds} (see also \cite{Herderschee:2021dez}), as well as Yangian invariants/plabic graphs~\cite{Mago:2020kmp, He:2020uhb,Mago:2020nuv}.

On the other hand, ${\cal N}=4$ SYM has proved to be an extremely fruitful laboratory for the study of Feynman integrals. For example, important ideas and powerful tools such as symbol and co-products~\cite{Goncharov:2010jf, Duhr:2011zq}, integrals with unit leading singularity and ${\rm d} \log$ forms~\cite{ArkaniHamed:2010gh} and even differential equations~\cite{Drummond:2010cz,Henn:2013pwa,Henn:2014qga}, have all more or less originated from the study in ${\cal N}=4$ but they have much wider applications. One of the most recent examples of this kind, which was motivated by~\cite{CaronHuot:2011ky}, is the so-called Wilson-loop ${\rm d}\log$ forms for a large class of Feynman integrals, based on the duality between amplitudes and Wilson loops in the theory~\cite{Alday:2007hr,Alday:2007he,Drummond:2007aua,Brandhuber:2007yx,Mason:2010yk,CaronHuot:2010ek}. Various ladder integrals, and {\it e.g.} the generic double-pentagon integrals for two-loop MHV and NMHV (component) amplitudes~\cite{He:2020uxy,He:2020lcu}, can be computed efficiently in this way, and we believe it to be closely related to the differential-equation method. 

Remarkably, the connection to cluster algebras extend to Feynman integrals as well: {\it e.g.} the symbol alphabet of six-point double-penta-ladder integral {\it etc.} was given by $A_3$ cluster algebra~\cite{Caron-Huot:2018dsv}, and the so-called cluster adjacency condition was observed for certain seven-point integrals in $E_6$~\cite{Drummond:2017ssj}. One can bootstrap Feynman integrals~\cite{Chicherin:2017dob, Henn:2018cdp} based on such knowledge (see also \cite{Dixon:2020bbt}). Very recently, the authors of~\cite{Chicherin:2020umh} have argued that cluster algebra structures appear for rather general Feynman integrals which go way beyond planar ${\cal N}=4$ SYM. They have provided strong evidence that four-point Feynman integrals with an off-shell leg is controlled by a $C_2$ cluster algebra, and found cluster-algebra alphabets for various one-loop integrals, as well as the general five-particle alphabet. A very natural question is how the alphabet may change as we go to higher loops for certain Feynman integrals: for six-point double-penta-ladder integrals, the alphabet stays as $A_3$ as mentioned~\cite{Caron-Huot:2018dsv}, and the main goal of the paper is to extend this to  more general cases.

\begin{figure} [htbp]
    \begin{tikzpicture}[baseline={([yshift=-.5ex]current bounding box.center)},scale=0.15]
                \draw[black,thick] (0,5)--(-5,5)--(-5,0)--(0,0)--cycle;
                \draw[black,thick] (1.93,5.52)--(0,5)--(0.52,6.93);
                \draw[black,thick] (1.93,-0.52)--(0,0)--(0.52,-1.93);
                \draw[thick,densely dashed] (-10,0) -- (-5,0);
                \draw[thick,densely dashed] (-10,5) -- (-5,5);
                \draw[black,thick] (-10,0)--(-10,5)--(-15,5)--(-15,0)--cycle;
                \draw[black,thick] (-16.93,5.52)--(-15,5)--(-15.52,6.93);
                \draw[black,thick] (-16.93,-0.52)--(-15,0)--(-15.52,-1.93);
                \filldraw[black] (1.93,6) node[anchor=west] {{$2$}};
                \filldraw[black] (0.52,6.93) node[anchor=south] {{$1$}};
                \filldraw[black] (1.93,-1) node[anchor=west] {{$3$}};
                \filldraw[black] (0.52,-1.93) node[anchor=north] {{$4$}};
                \filldraw[black] (-16.93,6) node[anchor=east] {{$7$}};
                \filldraw[black] (-15.52,6.93) node[anchor=south] {{$8$}};
                \filldraw[black] (-16.93,-1) node[anchor=east] {{$6$}};
                \filldraw[black] (-15.52,-1.93) node[anchor=north] {{$5$}};
                \filldraw[black] (-7.52,6.93) node[anchor=south] {{$x_1$}};
                \filldraw[black] (1.93,2.02) node[anchor=west] {{$x_3$}};
                \filldraw[black] (-7.52,-1.93) node[anchor=north] {{$x_5$}};
                \filldraw[black] (-16.93,2.02) node[anchor=east] {{$x_7$}};
            \end{tikzpicture}
\begin{tikzpicture}[baseline={([yshift=-.5ex]current bounding box.center)},scale=0.15]
                \draw[black,thick] (0,0)--(0,5)--(4.76,6.55)--(7.69,2.5)--(4.76,-1.55)--cycle;
                \draw[decorate, decoration=snake, segment length=12pt, segment amplitude=1.5pt, black,thick] (4.76,6.55)--(4.76,-1.55);
                \draw[black,thick] (9.43,1.5)--(7.69,2.5)--(9.43,3.5);
                \draw[black,thick] (4.76,6.55)--(5.37,8.45);
                \draw[black,thick] (4.76,-1.55)--(5.37,-3.45);
                \draw[black,thick] (0,5)--(-5,5)--(-5,0)--(0,0);
                \draw[thick,densely dashed] (-10,0) -- (-5,0);
                \draw[thick,densely dashed] (-10,5) -- (-5,5);
                \draw[black,thick] (-10,0)--(-10,5)--(-15,5)--(-15,0)--cycle;
                \draw[black,thick] (-16.93,5.52)--(-15,5)--(-15.52,6.93);
                \draw[black,thick] (-16.93,-0.52)--(-15,0)--(-15.52,-1.93);
                \filldraw[black] (5.37,8.45) node[anchor=south] {{$1$}};
                \filldraw[black] (5.37,-3.45) node[anchor=north] {{$4$}};
                \filldraw[black] (9.43,1.4) node[anchor=west] {{$3$}};
                \filldraw[black] (9.43,3.6) node[anchor=west] {{$2$}};
                \filldraw[black] (-16.93,5.52) node[anchor=east] {{$7$}};
                \filldraw[black] (-15.52,6.93) node[anchor=south] {{$8$}};
                \filldraw[black] (-16.93,-0.52) node[anchor=east] {{$6$}};
                \filldraw[black] (-15.52,-1.93) node[anchor=north] {{$5$}};
            \end{tikzpicture}
\begin{tikzpicture}[baseline={([yshift=-.5ex]current bounding box.center)},scale=0.15]
        \draw[black,thick] (0,0)--(0,5)--(4.76,6.55)--(7.69,2.5)--(4.76,-1.55)--cycle;
        \draw[black,thick] (-15,5)--(-19.76,6.55)--(-22.69,2.5)--(-19.76,-1.55)--(-15,0);
        \draw[decorate, decoration=snake, segment length=12pt, segment amplitude=2pt, black,thick] (4.76,6.55)--(4.76,-1.55);
        \draw[decorate, decoration=snake, segment length=12pt, segment amplitude=2pt, black,thick] (-19.76,6.55)--(-19.76,-1.55);
        \draw[black,thick] (9.43,1.5)--(7.69,2.5)--(9.43,3.5);
        \draw[black,thick] (4.76,6.55)--(5.37,8.45);
        \draw[black,thick] (4.76,-1.55)--(5.37,-3.45);
        \draw[black,thick] (0,5)--(-5,5)--(-5,0)--(0,0);
        \draw[black,thick,densely dashed] (-5,5)--(-10,5);
        \draw[black,thick,densely dashed] (-5,0)--(-10,0);
        \draw[black,thick] (-10,0)--(-10,5)--(-15,5)--(-15,0)--cycle;
        \draw[black,thick] (-19.76,6.55)--(-20.37,8.45);
        \draw[black,thick] (-19.76,-1.55)--(-20.37,-3.45);
        \draw[black,thick] (-22.69,2.5)--(-24.69,2.5);
        \filldraw[black] (-20.37,8.45) node[anchor=south] {{7}};
        \filldraw[black] (5.37,8.45) node[anchor=south] {{1}};
        \filldraw[black] (9.43,3.5) node[anchor=west] {{2}};
        \filldraw[black] (9.43,1.5) node[anchor=west] {{3}};
        \filldraw[black] (5.37,-3.45) node[anchor=north] {{4}};
        \filldraw[black] (-20.37,-3.45) node[anchor=north] {{5}};
        \filldraw[black] (-24.69,2.5) node[anchor=east] {{6}};
    \end{tikzpicture}
    \caption{The (eight-point) box-ladder, (eight-point) penta-box-ladder and (seven-point) double-penta-ladder integrals.}\label{fig1}
\end{figure}
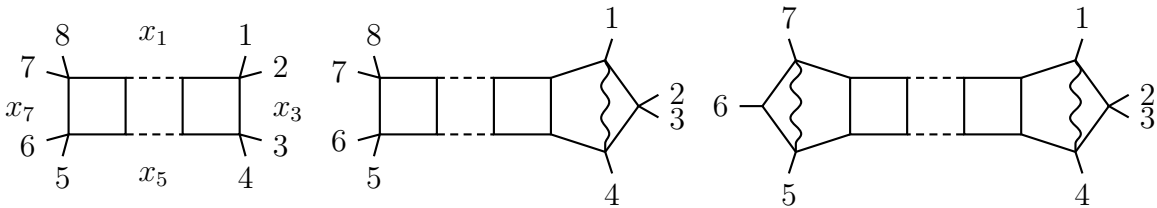

In this paper, we mainly show that (generic, eight-point) penta-box-ladder and (seven-point) double-penta-ladder have alphabets which correspond to cluster algebra $D_3\simeq A_3$ and $D_4$ respectively; as a toy example, we also present the trivial case of (eight-point) box-ladder which has alphabet $D_2\simeq A_1^2$. For the non-trivial ladders, we make the claim based on explicit calculations up to five loops (including all odd-weight cases in between). As shown in Fig.~\ref{fig1}, let us denote these three classes of integrals at $L$-loop as $I^{(L)}_{\rm b}(x_1, x_3, x_5, x_7)$, $I^{(L)}_{\rm pb} (x_1, x_2, x_4, x_5, x_7)$ and $I^{(L)}_{\rm dp}(x_1, x_2, x_4, x_5, x_6, x_7)$ which depend on $4$, $5$ and $6$ dual points, respectively. 

Before we give the precise definition of these classes of ladder integrals, let us first review the kinematics. Recall that the dual points are related to $n$ ordered, massless momenta by $p^\mu_i=x^\mu_{i{+}1}-x^\mu_{i}$, thus they form an null polygon with $n$ edges, labelled by external legs $i=1,2,\cdots, n$. It is convenient to introduce (supersymmetric) momentum twistors \cite{Hodges:2009hk}, ${\bf Z}:=Z^A_{i}$ with the ${\rm SL}(4)$ label $A=1,2,3,4$, defined as
\[
Z^A_{i}:=(\lambda_{i}^{\alpha},x_{i}^{\alpha\dot{\alpha}}\lambda_{i\alpha}).
\]
Each point $x_i$ in dual point space (vertex of the null polygon) corresponds to a line $(i{-}1 i)$ determined by two momentum twistors ${Z}_{i{-}1}$ and ${Z}_i$. Each loop momentum is represented by a point $y_{\ell}$ in the dual space, which also becomes a line/bi-twistor $\ell:= (AB)$ in twistor space. Squared-distance of two dual points then reads $x_{ij}^2:=\frac{\langle i{-}1 ij{-}1 j\rangle}{\langle {i{-}1i}\rangle\langle {j{-}1j}\rangle}$, where $\langle i{-}1 ij{-}1 j\rangle$ is the four bracket (the basic $SL(4)$ invariant) of four momentum twistors $\langle ijkl\rangle:=\epsilon_{a,b,c,d} Z_i^a Z_j^b Z_k^c Z_l^d$. Similarly we also have the definition $\langle \ell\,i{-}1 i\rangle:=\langle A B i{-}1 i\rangle=(x_\ell-x_i)^2 \langle A B \rangle \langle i{-}1 i\rangle$. 
 
The box-ladder integral, $I^{(L)}_{\rm b}$, which has no non-trivial numerator, is defined as (in terms of dual points and momentum twistors):

\vspace{-2.5ex}
{
\footnotesize
\begin{equation}
        \begin{split}
            I^{(L)}_{\rm b}(x_1,x_3,x_5,x_7) &= \displaystyle\int\frac{{\rm d}^{4L}y}{(i\pi^2)^{L}}\dfrac{x_{37}^2(x_{15}^2)^L}{(y_1{-}x_3)^2\left[\displaystyle\prod_{a=1}^{L{-}1}y_{a{-}1a}^2 (y_a{-}x_1)^2(y_a{-}x_5)^2\right](y_L{-}x_1)^2(y_L{-}x_5)^2 (y_L{-}x_7)^2}            \\
            &= \displaystyle\int{\rm d}^{4L}\ell\dfrac{\langle2367\rangle\langle8145\rangle^L}{\langle\ell_123\rangle\left[\displaystyle\prod_{a=1}^{L{-}1}\langle\ell_{a{-}1}\ell_a\rangle\langle\ell_a81\rangle\langle\ell_a45\rangle\right]\langle\ell_L81\rangle\langle\ell_L45\rangle\langle\ell_{L}67\rangle}.
        \end{split}
    \end{equation} 
    }%
where we have denoted loop variables from the right-most to the left by dual points $\{y_1,\cdots, y_L\}$ (or bi-twistors $\{\ell_1,\dots,\ell_L\}$), and $y^2_{ab}:=(y_a-y_b)^2$. 

For $I^{(L)}_{\rm pb}$ and $I^{(L)}_{\rm dp}$, it is more convenient to directly write them using momentum twistors, especially for the ``wavy-line" numerators~\cite{Drummond:2010cz,ArkaniHamed:2010gh}: 
\begin{equation}
        \begin{split}
            I^{(L)}_{\rm pb}(x_1, x_2, x_4, x_5, x_7) 
            &= \displaystyle\int{\rm d}^{4L}\ell\dfrac{\langle\ell_1\bar 1\cap\bar 4\rangle\langle 1467\rangle\langle 8145\rangle^{L-1}}{\llangle\ell_11\rrangle\llangle\ell_14\rrangle\left[\displaystyle\prod_{a=2}^{L}\langle\ell_{a{-}1}\ell_a\rangle\langle\ell_a81\rangle\langle\ell_a45\rangle\right]\langle\ell_{L}67\rangle},
        \end{split}
    \end{equation} 
\begin{equation}\label{doublepentaladder7}
\begin{split}
        I^{(L)}_{\rm dp}(x_1, x_2, x_4, x_5, x_6, x_7)
        &=\displaystyle\int{\rm d}^{4L}\ell\dfrac{\langle\ell_1\bar 1\cap\bar 4\rangle\langle\ell_L \bar 5\cap\bar 7\rangle\langle 1457\rangle^{L-1}}{\llangle\ell_11\rrangle\llangle\ell_14\rrangle\left[\displaystyle\prod_{a=2}^{L}\langle\ell_{a{-}1}\ell_a\rangle\langle\ell_a71\rangle\langle\ell_a45\rangle\right]\llangle\ell_L5\rrangle\llangle\ell_L7\rrangle}
\end{split}
\end{equation}
where we have introduced the shorthand notation $\llangle\ell i\rrangle:=\langle\ell i{-}1i\rangle\langle\ell ii{+}1\rangle$. Note that we can alternatively adopt the ``dashed-line" numerator $\langle \ell i j\rangle$ which is proportional to our $\langle \ell \bar{i} \cap \bar{j}\rangle$ (for $I_{\rm dp}$ we need to replace both wavy-lines by dashed-lines to have a pure function). The alphabet of $I_{\rm pb}$ and $I_{\rm dp}$ is not affected by such a parity conjugation. 
  
All these integrals can be evaluated relatively straightforwardly: in addition to the well-known box ladder-integrals, in~\cite{He:2020uxy} we have proposed a recursive formula for the other two classes of integrals, in terms of the so-called Wilson-loop ${\rm d}\log$ representation; explicitly the chiral-pentagon on the right-end can be written as two-fold ${\rm d}\log$ integrals of a $(L{-}1)$-loop integral where the right-end is again a pentagon with deformed legs, {\it e.g.} for $I_{\rm pb}^{(L)}$ we have the following recursion with $X_1=Z_8-\tau_X Z_2$ and $Y_1=Z_3-\tau_Y Z_5$ (more details can be found in~\cite{He:2020uxy} and below):
\begin{equation}\label{pentarecursion}
        \begin{split}
            I^{(L)}_{\rm pb}(x_1, x_2, x_4, x_5, x_7) &=  \displaystyle\int {\rm d} \log\langle 814Y_1\rangle \,{\rm d}\log\frac{\langle 1X_14Y_1\rangle}{\tau_{X_1}}  \begin{tikzpicture}[baseline={([yshift=-.5ex]current bounding box.center)},scale=0.18]
                \draw[black,thick] (0,0)--(0,5)--(4.76,6.55)--(7.69,2.5)--(4.76,-1.55)--cycle;
                \draw[decorate, decoration=snake, segment length=12pt, segment amplitude=2pt, black,thick] (4.76,6.55)--(4.76,-1.55);
                \draw[black,thick] (9.43,1.5)--(7.69,2.5)--(9.43,3.5);
                \filldraw[black] (9.19,2.5) circle [radius=2pt];
                \filldraw[black] (9.1,1.99) circle [radius=2pt];
                \filldraw[black] (9.1,3.01) circle [radius=2pt];
                \draw[black,thick] (4.76,6.55)--(5.37,8.45);
                \draw[black,thick] (4.76,-1.55)--(5.37,-3.45);
                \draw[thick,densely dashed] (-5,0) -- (0,0);
                \draw[thick,densely dashed] (-5,5) -- (0,5);
                \draw[black,thick] (-5,0)--(-5,5)--(-10,5)--(-10,0)--cycle;
                \draw[black,thick] (-11.93,5.52)--(-10,5)--(-10.52,6.93);
                \draw[black,thick] (-11.93,-0.52)--(-10,0)--(-10.52,-1.93);
                \filldraw[black] (-11.93,5.52) node[anchor=east] {{$7$}};
                \filldraw[black] (-10.52,6.93) node[anchor=south] {{$8$}};
                \filldraw[black] (-11.93,-0.52) node[anchor=east] {{$6$}};
                \filldraw[black] (-10.52,-1.93) node[anchor=north] {{$5$}};
                \filldraw[black] (5.37,8.45) node[anchor=south west] {{$1$}};
                \filldraw[black] (5.37,-3.45) node[anchor=north west] {{$4$}};
\filldraw[black] (9.43,3.5) node[anchor=south west] {{$X_1$}};
                \filldraw[black] (9.43,1.5) node[anchor=north west] {{$Y_1$}};
            \end{tikzpicture} \\[-3ex]
&= \displaystyle\int {\rm d} \log\langle 814Y_1\rangle {\rm d}\log\frac{\langle 1X_14Y_1\rangle}{\tau_{X_1}} \tilde I^{(L-1)}_{\rm pb}.
        \end{split}
 \end{equation}
 
After we finished the first version of the manuscript, we noticed that our double-penta-ladder integral $I_{\rm dp}$ has been evaluated up to $L=3$ (denoted as {\bf heptagon A}) in~\cite{Bourjaily:2018aeq}. It is interesting to note that the authors of \cite{Bourjaily:2018aeq} have also evaluated examples of other double-penta-ladder integrals for heptagon and octagon cases using Feynman parametrization. 
We will discuss such integrals in detail in section \ref{sec3}, and let us briefly summarize the main result here, which shows the remarkable universality of cluster algebra structures for these Feynman integrals. For our purposes, the most generic case is the nine-point double-penta-ladder integrals:
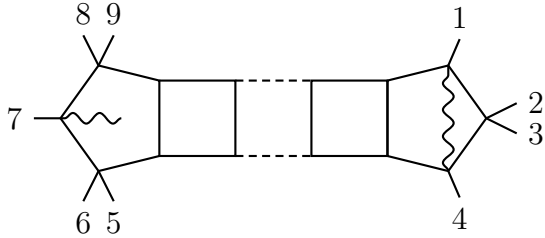
\[
\begin{tikzpicture}[scale=0.5]
\draw[thick] (-1,2) -- (-2.6,2.4) -- (-3.6,1) -- (-2.6,-0.4) -- (-1,0)--cycle;
\draw[thick] (5,2) -- (6.6,2.4) -- (7.6,1) -- (6.6,-0.4) -- (5,0)--cycle;
\draw[thick] (-2.6,2.4) -- (-2.2,3.2) node[above]{$9$};
\draw[thick] (-2.6,2.4) -- (-3,3.2) node[above]{$8$};
\draw[thick] (-3.6,1) -- (-4.3,1) node[left]{$7$};
\draw[thick] (-2.6,-0.4) -- (-3,-1.2) node[below]{$6$};
\draw[thick] (-2.6,-0.4) -- (-2.2,-1.2) node[below]{$5$};
\draw[thick] (6.6,-0.4) -- (6.9,-1.1) node[below]{$4$};
\draw[thick] (7.6,1) -- (8.4,0.6) node[right]{$3$};
\draw[thick] (7.6,1) -- (8.4,1.4) node[right]{$2$};
\draw[thick] (6.6,2.4) -- (6.9,3.1) node[above]{$1$};
\draw[thick] (-1,2) -- (1,2) -- (1,0) -- (-1,0);
\draw[thick,densely dashed] (1,2) -- (3,2);
\draw[thick,densely dashed] (1,0) -- (3,0);
\draw[thick] (3,2) -- (5,2) -- (5,0) -- (3,0) -- cycle;
\draw[decorate, decoration={snake, segment length=10pt, amplitude=2pt}, black,thick] (6.6,-0.4)--(6.6,2.4);
\draw[decorate, decoration={snake, segment length=10pt, amplitude=2pt}, black,thick] (-2,1)--(-3.6,1);
\end{tikzpicture}
\]
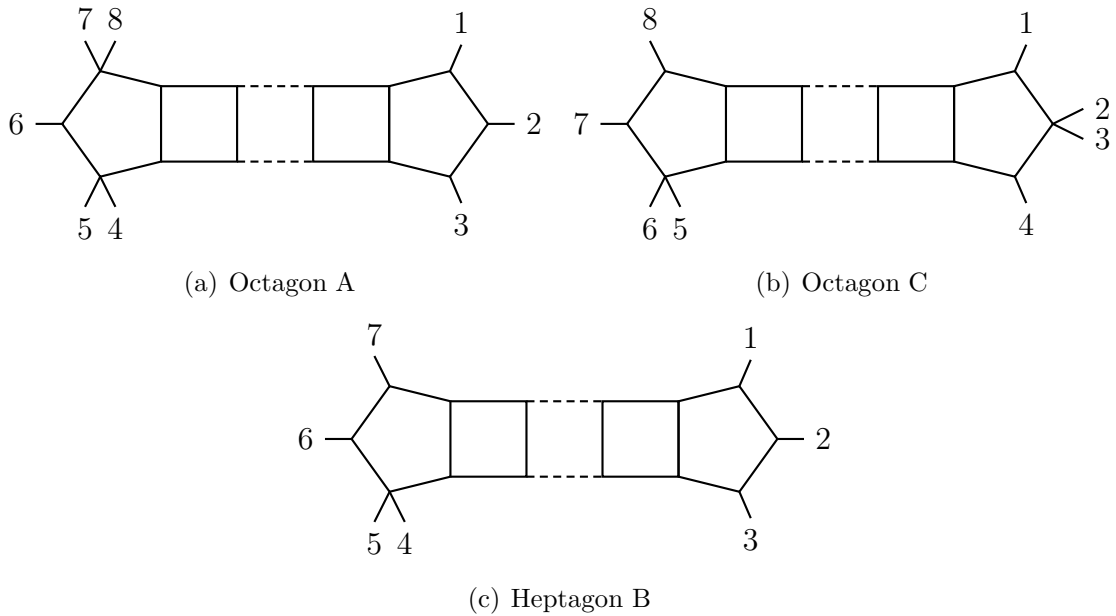
\begin{figure}[htbp]
\centering
\subfigure[Octagon A]{
\begin{tikzpicture}[scale=0.5]
\draw[thick] (-1,2) -- (-2.6,2.4) -- (-3.6,1) -- (-2.6,-0.4) -- (-1,0)--cycle;
\draw[thick] (5,2) -- (6.6,2.4) -- (7.6,1) -- (6.6,-0.4) -- (5,0)--cycle;
\draw[thick] (-2.6,2.4) -- (-2.2,3.2) node[above]{$8$};
\draw[thick] (-2.6,2.4) -- (-3,3.2) node[above]{$7$};
\draw[thick] (-3.6,1) -- (-4.3,1) node[left]{$6$};
\draw[thick] (-2.6,-0.4) -- (-3,-1.2) node[below]{$5$};
\draw[thick] (-2.6,-0.4) -- (-2.2,-1.2) node[below]{$4$};
\draw[thick] (6.6,-0.4) -- (6.9,-1.1) node[below]{$3$};
\draw[thick] (7.6,1) -- (8.3,1) node[right]{$2$};
\draw[thick] (6.6,2.4) -- (6.9,3.1) node[above]{$1$};
\draw[thick] (-1,2) -- (1,2) -- (1,0) -- (-1,0);
\draw[thick,densely dashed] (1,2) -- (3,2);
\draw[thick,densely dashed] (1,0) -- (3,0);
\draw[thick] (3,2) -- (5,2) -- (5,0) -- (3,0) -- cycle;
\end{tikzpicture}}%
\subfigure[Octagon C]{
\begin{tikzpicture}[scale=0.5]
\draw[thick] (-1,2) -- (-2.6,2.4) -- (-3.6,1) -- (-2.6,-0.4) -- (-1,0)--cycle;
\draw[thick] (5,2) -- (6.6,2.4) -- (7.6,1) -- (6.6,-0.4) -- (5,0)--cycle;
\draw[thick] (-2.6,2.4) -- (-3,3.2) node[above]{$8$};
\draw[thick] (-3.6,1) -- (-4.3,1) node[left]{$7$};
\draw[thick] (-2.6,-0.4) -- (-3,-1.2) node[below]{$6$};
\draw[thick] (-2.6,-0.4) -- (-2.2,-1.2) node[below]{$5$};
\draw[thick] (6.6,-0.4) -- (6.9,-1.1) node[below]{$4$};
\draw[thick] (7.6,1) -- (8.4,0.6) node[right]{$3$};
\draw[thick] (7.6,1) -- (8.4,1.4) node[right]{$2$};
\draw[thick] (6.6,2.4) -- (6.9,3.1) node[above]{$1$};
\draw[thick] (-1,2) -- (1,2) -- (1,0) -- (-1,0);
\draw[thick,densely dashed] (1,2) -- (3,2);
\draw[thick,densely dashed] (1,0) -- (3,0);
\draw[thick] (3,2) -- (5,2) -- (5,0) -- (3,0) -- cycle;
\end{tikzpicture}
}
\subfigure[Heptagon B]{
\begin{tikzpicture}[scale=0.5]
\draw[thick] (-1,2) -- (-2.6,2.4) -- (-3.6,1) -- (-2.6,-0.4) -- (-1,0)--cycle;
\draw[thick] (5,2) -- (6.6,2.4) -- (7.6,1) -- (6.6,-0.4) -- (5,0)--cycle;
\draw[thick] (-2.6,2.4) -- (-3,3.2) node[above]{$7$};
\draw[thick] (-3.6,1) -- (-4.3,1) node[left]{$6$};
\draw[thick] (-2.6,-0.4) -- (-3,-1.2) node[below]{$5$};
\draw[thick] (-2.6,-0.4) -- (-2.2,-1.2) node[below]{$4$};
\draw[thick] (6.6,-0.4) -- (6.9,-1.1) node[below]{$3$};
\draw[thick] (7.6,1) -- (8.3,1) node[right]{$2$};
\draw[thick] (6.6,2.4) -- (6.9,3.1) node[above]{$1$};
\draw[thick] (-1,2) -- (1,2) -- (1,0) -- (-1,0);
\draw[thick,densely dashed] (1,2) -- (3,2);
\draw[thick,densely dashed] (1,0) -- (3,0);
\draw[thick] (3,2) -- (5,2) -- (5,0) -- (3,0) -- cycle;
\end{tikzpicture}}
\caption{Degenerations of nine-point double-penta-ladder}\label{fig2}
\end{figure}%
and we remark that there are several natural choices of numerators (``wavy lines") which make the integral pure, and we write two of them explicitly in sec.~\ref{sec3} (see~\cite{Bourjaily:2018aeq}). We conjecture that the alphabet of these nine-point integrals (independent of the choice of numerator) is given by (a subset of) cluster algebra $D_6$. We can take collinear limit $3\to 2$ (and relabel $i{+}1 \to i$ for $i=3,\cdots, 8$) to obtain eight-point integral that has been denoted as {\bf octagon A}) (and evaluated to $L=4$) in~\cite{Bourjaily:2018aeq}. Note that the kinematics of our nine-point case is the same as three-mass-easy hexagon~\cite{DelDuca:2011wh}, and in the collinear limit, it becomes two-mass-easy case. There is no smooth limit in $9\to 8$ or $6\to 5$ (for reasons similar to that in~\cite{Bourjaily:2018aeq}), but one can define such eight-point double-penta-ladder integrals where we have two massive corners with legs $2,3$ and $5,6$ (we denote it as {\bf octagon C}). We can similarly define seven-point integrals which are different than our $I_{\rm dp}$ ({\bf heptagon A}) above, such as {\bf heptagon B} in~\cite{Bourjaily:2018aeq} (all these integrals are shown in figure.~\ref{fig2} without specifying numerators). We conjecture that the alphabet of any octagon integrals is given by $D_5$ cluster algebra and that of any heptagon ones is given by $D_4$ cluster algebra. It is remarkable that the alphabet for these integrals seem to be independent of any detail: not only the numerators but also different propagator structures (see the comparison of two types of octagons and heptagons). It is tempting to say that we can simply associate the three-mass-easy, two-mass-easy and one-mass hexagon kinematics (for $n=9,8,7$) with cluster algebras $D_6$, $D_5$ and $D_4$, respectively!

Such integrals evaluate to (linear combinations of) multiple polylogarithms, and there is a well-known Hopf algebra structure~\cite{Goncharov:2005sla}, which has led to the notion of \textit{symbol}~\cite{Goncharov:2010jf, Duhr:2011zq}. For any multiple polylogarithm $G^{(w)}$ whose differential reads
\[
dG^{(w)}=\sum_i G^{(w-1)}_i d\log x_i,
\]
where $w$ is called the weight of the polylogarithm and $\{G^{(w-1)}_i\}$ are polylogarithms of lower weight $w-1$, its symbol of $G^{(w)}$ is defined by
\[
\mathcal{S}(G^{(w)}):=\sum_i \mathcal{S}(G^{(w-1)}_i)\otimes x_i,
\]
and $G^{(0)}:=1$. For example,
\[
\mathcal{S}(\log(x))=x,\quad 
\mathcal{S}(\operatorname{Li}_2(x))=-\,(1-x)\otimes x.
\]
Therefore, the symbol of a polylogarithm of weight $w$ is a tensor of length $w$, whose entries are called \textit{letters}. The collection of all letters is called the \textit{alphabet}.

Given our recursion for the ladder integrals above, we can directly read off the symbol by the following rules~\cite{CaronHuot:2011kk}: Suppose we have an integral
\[
\int_a^b {\rm d}\log(t+c)\, (F(t)\otimes w(t)),
\]
where $F(t)\otimes w(t)$ is a integrable, linear reducible symbol in $t$, {\it i.e.} its entries are products of powers of linear polynomials in $t$, and $w(t)$ is the last entry. The total differential of this integral is the sum of the following two parts:
\begin{compactenum}[\quad (1)]
\item the contribution from endpoints:
\[
    {\rm d}\log(t+c)(F(t)\otimes w(t))|_{t=a}^{t=b}=(F(t)\otimes w(t)\otimes (t+c))|_{t=a}^{t=b},
\]
\item contributions from the last entry: for a term where $w(t)$ is a constant,
\[
\left(\int_a^b {\rm d}\log(t+c)\, F(t)\right){\rm d}\log w=\left(\int_a^b {\rm d}\log(t+c)\, F(t)\right)\otimes w,
\]
and for a term where $w(t)=t+d$,
\[
\left(\int_a^b {\rm d}\log \frac{t+c}{t+d}\, F(t)\right){\rm d}\log (c-d)
=\left(\int_a^b {\rm d}\log \frac{t+c}{t+d}\, F(t)\right)\otimes (c-d).
\]
\end{compactenum}
Then we can recursively compute the symbol of the lower-weight integrals and obtain the symbol of the iterated integral of ${\rm d} \log$ forms. 

We emphasize that it is also straightforward to compute the functions rather than the symbol, but to get the alphabet and other symbolic information we still need to take the symbol map. We have computed $I_{\rm pb}^{(L)}$ up to $L=5$ and $I_{\rm dp}^{(L)}$ up to $L=4$, as linear combinations of multiple polylogs~\footnote{Such computations can be done using our code on {\bf Mathematica} easily, {\it e.g.} the five-loop computation takes a few minutes on a laptop. They can also be done with {\it e.g.} HyperInt~\cite{Panzer:2014caa}.}. We find that the functions are actually quite nice at least using ``good variables" motivated by cluster algebras (see below), {\it e.g.} from $L=2$ to $L=3$, the number of mutliple polylogs in the answer grows from a dozen to a few hundreds at most. As an illustration we present a rather compact expression for $I_{\rm pb}^{(L)}$ at $L=3$.





\subsection{Review of cluster algebra and cluster configuration space}

Cluster algebras~\cite{fomin2002cluster,fomin2003cluster,berenstein2005cluster,fomin2007cluster} are commutative algebras with a particular set of generators ${\cal A}_i$, known as the cluster ${\cal A}$-coordinates; they are grouped into {\it clusters} which are subsets of rank $d$. From an initial cluster, one can construct all the ${\cal A}$-coordinates by {\it mutations} acting on ${\cal A}$'s (the so-called frozen coordinates or coefficients can also be included, which do not mutate). Alternatively one can define cluster ${\cal X}$ coordinates which are given by monomials of ${\cal A}$'s.  

There is a natural space of polylogarithm functions associated with a cluster algebra, given a set of cluster-${\cal A}$ or (${\cal X}$) coordinates. A cluster function $F^{(w)}$ \cite{Golden:2014xqa,Parker:2015cia} of transcendental weight $w$ is defined such that its differential has the form
\[
d F^{(w)}=\sum_i F_i^{(w-1)} d \log {\cal A}_i
\]
where $F^{(w-1)}_i$ are cluster functions of transcendental weight $w-1$ and ${\cal A}_i$ are cluster-${\cal A}$ coordinates. We see that if a multiple polylogarithm is a cluster function, then the alphabet can be identified with the corresponding cluster algebra. 

For the purpose of this paper, it suffices to know that all finite-type cluster algebras, {\it i.e.} those with finite number of $A$-cluster coordinates (the dimension of the cluster algebra, denoted as $N$), has been classified in terms of Dynkin diagrams. There are series $A_d, B_d, C_d, D_d$ and exceptional cases $E_6, E_7, E_8, F_4, G_2$. To identify an alphabet with certain finite-type cluster algebra, it is convenient to parametrize the coordinates of the latter in a nice way. For example, for type $A_d$ and $D_d$, we have the following $N=d(d{+}3)/2$ and $N=d^2$ letters respectively~\cite{Chicherin:2020umh}:

\vspace{-2.5ex}
{\footnotesize \begin{align}\label{AD}
& \Phi_{A_d}=\bigcup_{i=1}^d \{z_i, 1+z_i\} \cup \bigcup_{i<j}^d \{z_i-z_j\},\\
& \Phi_{D_d}=\bigcup_{i=1}^d \{z_i, 1{+}z_i\} \cup \bigcup_{i=1}^{d-2}\{z_i{+} z_{d-1} z_d, z_i{-}z_{d-1}, z_i{-}z_d\} \cup \bigcup_{i<j}^{d-2} \{z_i{-}z_j, z_i{-}z_j{-}z_i z_j {+} z_i (z_{d-1}{+}z_d){-}z_{d-1} z_d\} \nonumber
\end{align}}

In other words, once we find an alphabet which can be written as a collection of $N$ polynomials (of $d$ variables), the remaining task would be to look for some birational change of variables such that they become multiplicative combinations of letters in {\it e.g.} $\Phi_{A_d}$ or $\Phi_{D_d}$ (or a subset of them).

On the other hand, without any smart parametrization, there is a totally gauge-invariant way for describing any finite-type cluster algebra, known as the cluster configuration spaces~\cite{Arkani-Hamed:2019plo,Arkani-Hamed:2020tuz}. One can simply represent $\Phi$ by $N$ variables called ${\bf u}$ variables, $\{{\bf u}_\alpha | \alpha=1,2, \cdots, N\}$ in bijection with ${\cal A}$-coordinates, which satisfy $N$ constraints known as the ${\bf u}$ equations (for $\alpha=1,2,\cdots, N$)
\[1-{\bf u}_\alpha=\prod_{\beta=1}^N {\bf u}_\beta^{\beta| \alpha}\,,\]
where $\beta|\alpha$ are integers known as {\it compatibility degrees}~\cite{Arkani-Hamed:2019plo}. It is remarkable that the ${\bf u}$ equations are consistent and give a $d$-dimensional solution space we call {\it cluster configuration space}. For example, for type $A_d$ we have $N=d(d{+}3)/2$ ${\bf u}$ variables (one for each diagonal of $(d{+}3)$-gon):
\begin{equation}\label{uAn}
\Phi_{A_d}=\{{\bf u}_{ij} | 1\leq i<j \leq n, i\neq j-1\}
\end{equation}
which satisfy $N$ equations of the form
\begin{equation}\label{ueqAn}
1-{\bf u}_{i,j}=\prod_{(k l)\, {\rm  cross }\, (i j)} {\bf u}_{k,l}
\end{equation}
where on the RHS we have the product of all ${\bf u}_{k,l}$ with $(k,l)$ crossing $(i,j)$ (or ${\bf u}_{k,l}$ incompatible with ${\bf u}_{i,j}$, with compatibility $1$). Similarly for $D_d$ we have $N=d^2$ variables which we denote as 
\begin{equation}\label{uDn}
\Phi_{D_d}=\{{\bf u}_i, \tilde{\bf u}_i | 1\leq i \leq d\} \cup \{ {\bf u}_{i,j} | 1\leq  i\neq j, j{+}1  \leq d\}
\end{equation}
which satisfy ${\bf u}$ equations as explicitly given in~\cite{Arkani-Hamed:2019plo, Arkani-Hamed:2020tuz}. 

Note that such a configuration space can be viewed as a ``binary geometry" for the corresponding cluster polytope (or generalized associahedra). If we ask all ${\bf u}$ to be positive, we have $\{0<{\bf u}_\alpha<1 | \alpha=1,\cdots, N\}$, which cuts out a ``curvy" cluster polytope. Each of the $N$ boundaries of the space is reached by exactly one ${\bf u}_\alpha \to 0$, and the ${\bf u}$ equations force all incompatible ({\it i.e.} those with $\beta|\alpha>0$) ${\bf u}_\beta \to 1$, and we obtain the configuration space of the corresponding {\it sub-algebra}, which factorizes according to which node of the Dynkin diagram we remove. Note that though the complex configuration space is no longer a polytope, we still have such boundary structures exactly as any $u_{\alpha} \to 0$. 

How do we find the ${\bf u}$ variables given an alphabet of $N$ polynomials? This has been proposed in~\cite{Arkani-Hamed:2019mrd}, and the basic idea is to use any {\it positive} parametrization such that the $N$ polynomials can be expressed as {\it subtraction-free} Laurant polynomials of positive coordinates $x_1, \cdots, x_d$, which we denote as $p_I(\{x\})$ for $I=1,\cdots, N$; then we study the so-called {\it stringy canonical forms}~\cite{Arkani-Hamed:2019mrd}:
\begin{equation}
I_{\{p\}}(\{s\})=(\alpha')^d \int_{\mathbb{R}_{>0}^d} \prod_{i=1}^d d\log x_i \prod_{I=1}^N p_I(\{x\})^{\alpha' s_I}\,,
\end{equation}
where we integrate $\prod_i {\rm d}\log x_i$ in the positive domain, and we take polynomials $p_I$ (to the power $\alpha' s_I$ as ``regulators" for potential divergences. The domain of the convergence for $I_{\{p\}}(\{s\})$ is given by a {\it polytope} in the exponent space, which is defined as the the Minkowski sum of Newton polytopes of $p_I$'s~\footnote{The $\alpha'\to 0$ limit of the integral itself is given by the canonical form of the polytope (hence the name).}. One can generally define ${\bf u}$ variables (and configuration spaces) for such an integral following the procedure in~\cite{Arkani-Hamed:2019mrd}, and quite beautifully all finite-type cluster algebras belong to a special case that the corresponding polytope has exactly $N$ facets and it is cut out by $X_{\alpha} (\{s\}) \geq 0$ for $\alpha=1, \cdots, N$, known as the ABHY realization of cluster polytopes; for type $A,B,C,D$, the canonical form of these polytopes, or $\alpha'\to 0$ limit of the integrals, have nice interpretation as planar $\phi^3$ amplitudes through one-loop~\cite{Arkani-Hamed:2017mur,Arkani-Hamed:2019vag}. We can recombine the exponents into $X_{\alpha}$'s and the regulator becomes $\prod_{\alpha=1}^N {\bf u}_{\alpha}^{\alpha' X_{\alpha}}$, where the polynomials combine into exactly the $N$ ${\bf u}_{\alpha}$ variables! Thus the ${\bf u}$ variables can be obtained by a Minkowski-sum calculation given any positive parametrization.

For example, by using any positive parametrization of polynomials in $\Phi_{A_d}$, we recognize $I_{\{p\}}$ as the usual $(d{+}3)$-point open-string integral, and the Minkowski sum gives exactly the ABHY associahedron in the kinematic (Mandelstam) space; the ${\bf u}_{ij} $ variables are then dihedral coordinates of ${\cal M}_{0, d{+}3}$~\cite{brown2009multiple}, given by cross-ratios of the world-sheet coordinates $z_i$'s (with three additional fixed at $(0, -1, \infty)$). 

\section{Cluster algebras for three classes of ladder integrals}
\subsection{$D_2\simeq A_1^2$ for box-ladder integrals}
Let us start with the well-known ladder integral, $I^{(L)}_{\rm b}(x_1, x_3, x_5, x_7)$ which depends on two cross-ratios and it is convenient to introduce $z$ and $\bar{z}$ variables defined by
\begin{equation}\label{zzb}
    \frac{z \bar{z}}{(1-z)(1-\bar{z})}=\frac{x_{13}^2 x_{57}^2}{x_{15}^2 x_{37}^2}\,,\quad 
   \frac{1}{(1-z)(1-\bar{z})}=\frac{x_{17}^2 x_{35}^2}{x_{15}^2 x_{37}^2}\,.
\end{equation}
The ladder integrals have been evaluated in~\cite{ussyukina1993exact}, and one way to do so is by solving differential equations they satisfy, which nicely relate $I^{(L)}_{\rm b}$ to $I^{(L{-}1)}_{\rm b}$. Recall that the integral has a natural overall normalization factor $I^{(L)}_{\rm b}:=f^{(L)}/(z-{\bar z})$ such that $f^{(L)}$ becomes pure function of weight $2L$. The functions  $f^{(L)}$ satisfy second-order differential equations:
\begin{equation}
   z  \partial_z \bar{z} \partial_{\bar{z}} f^{(L)}(z, \bar{z})=f^{(L{-}1)} (z, \bar{z}),
\end{equation}
with ``tree-level source" $f^{(0)}=\frac {z \bar{z}}{(1-z)(1-\bar{z})}$. There is a closed-form expression for $f^{(L)}$ from solving the differential equations:
\begin{equation}\label{fL}
f^{(L)}=\sum_{m=L}^{2L} \frac{m!\ (\log(-z \bar{z}))^{2L-m}}{L!\ (m-L)!\ (2L-m)!}\left({\rm Li}_m (z)-{\rm Li}_m (\bar{z}) \right)
\end{equation}
which is a single-valued, analytic function of $z$ (In Euclidean signature, $z$ and $\bar{z}$ are complex conjugates to each other). From \eqref{fL} it is obvious that the alphabet of the symbol consists of $4$ letters, $\{z, \bar{z}, 1-z, 1-\bar{z}\}$, which we can immediately identify as that of $D_2 \simeq A_1^2$. 

We are mostly interested in {\it positive} external kinematics, where momentum twistors ${\bf Z} \in G_+(4,n)$, and it is easy to see that for such kinematics we have $z<0$ and $\bar{z}<0$. To relate this alphabet to the {\it positive} ${\bf u}$ variables of $D_2$, we make the change of variables ${\bf u}:=z/(z-1)$ and ${\bf \bar{u}}:=\bar{z}/(\bar{z}-1)$. The $D_2$ alphabet can be alternatively written in these variables~\footnote{Our convention follows that of~\cite{Drummond:2010cz}, and differs from \cite{He:2020lcu} where the $z$ and $\bar{z}$ would be the variables ${\bf u}$ and ${\bf \bar{u}}$ here.}:
\begin{equation}
{\bf A}[I_{\rm b}^{(L)}]=\{{\bf u}, 1-{\bf u}, {\bf \bar{u}}, 1-{\bf \bar{u}}\}
\end{equation}
where all the letters are {\it positive} (between $0$ and $1$); this is the ${\bf u}$ space of $D_2$ which is literally a quadralateral. 

The variables $z, {\bar z}$ (or equivalently ${\bf u}, {\bf \bar{u}}$) are algebraic functions of momentum twistors, since they are two roots of the above quadratic equation. Without loss of generality for this specific problem, we can ``rationalize" the square root explicitly by reducing the kinematics to two dimensions~\cite{Caron-Huot:2013vda}. Recall that when external kinematics lie in two-dimensional subspace, the polygon can only have even number of edges (which we denote as $2n$) and take a zigzag shape: edges with even and odd labels go along two light-like directions respectively. It is convenient to reduce momentum twistors as
$Z_{2i-1}=(\lambda_{2i-1}^1, 0, \lambda_{2i-1}^2,0)$ and $Z_{2i}=(0, \tilde\lambda_{2i}^1, 0, \tilde\lambda_{2i}^2)$, which reduces the conformal group ${\rm SL}(4)$ to ${\rm SL}(2)\times {\rm SL}(2)$. The kinematics are encoded in even and odd ${\rm SL}(2)$ invariants $\langle i\,j\rangle$ (or $[i\,j]$) for odd (or even) $i,j$, which are in fact one-dimensional distances along odd (or even) direction. 

Any cross-ratio factorizes into the product of an even and an odd cross-ratio, {\it e.g.} for $i,j,k,l$ all even, we have $\frac{x_{ij}^2 x_{kl}^2}{x_{ik}^2 x_{jl}^2} \to u^{\rm 2d}_{i-1, j-1, k-1, l-1} u^{\rm 2d}_{i,j,k,l}$ where $u^{\rm 2d}_{i,j,k,l}:=\frac{[i\,j][k,l]}{[i\,k][j\,l]}$ denote familiar cross-ratios of $A_{n{-}3} \sim G(2,n)/T$  in the even sector (and similarly an $A_{n{-}3}$ in the odd sector). Now for the box-ladder with $2n=8$, we see that the $2d$ kinematics naturally require two $A_1$'s (for even and odd sectors), and we the square root disappear to give~\cite{Caron-Huot:2013vda}
\[
{\bf u}=u^{\rm 2d}_{1,3,5,7}\,,\quad {\bf \bar{u}}=u^{\rm 2d}_{2,4,6,8}
\]
and $1-{\bf u}=u^{\rm 2d}_{3,5,7,1}$ {\it etc.}. Thus we see that the ${\bf u}$ variables are literally the ${\bf u}$ variables for the two $A_1\sim G(2,4)/T$. 

Moreover, it is trivial to see that we have $A_1$ sub-algebras of $D_2$ which can be reached when any of the ${\bf u}$ variables goes to zero. This is well known since {\it e.g.} at one-loop, as ${\bf u} \to 0$ the box function diverges, but we can look at the ``finite part" ${\rm Li}_2 (1-{\bf \bar{u}})$ which has the $A_1$ alphabet $\{ {\bf \bar u}, 1-{\bf \bar u} \}$. Although this particular integral $I_{\rm b}^{(L)}$ diverges at any ``boundary" of the $D_2$, more generic $D_2$ cluster functions can have such $A_1$ functions in these limits.  

\subsection{$D_3 \simeq A_3$ for penta-ladder integrals}

Next we consider a more non-trivial example, $I^{(L)}_{\rm pb} (x_1, x_2, x_4, x_5, x_7)$, which depends on $3$ cross-ratios defined as
\begin{equation}
u=\frac{x_{17}^2 x_{25}^2}{x_{15}^2 x_{27}^2}\,,\quad v=\frac{x_{14}^2 x_{57}^2}{x_{15}^2 x_{47}^2}\,, \quad w=\frac{x_{15}^2 x_{24}^2}{x_{14}^2 x_{25}^2}\,.
\end{equation}
As shown in~\cite{He:2020uxy}, we obtain a two-step recursion relation for $I_{\rm pl}(u,v,w)$:
\begin{equation}\label{penta}
\begin{split}
I_{\rm pb}^{(L{+}\frac12)}(u,v,w)&=\int_0^\infty {\rm d} \log\frac{t+1}{t}\ I_{\rm pb}^{(L)}\left(\frac{u(t+w)}{t+u w},v,\frac{w(t+1)}{t+w}\right)\\
I_{\rm pb}^{(L{+}1)}(u,v,w)&=\int_0^\infty {\rm d} \log(s+1)\ I_{\rm pb}^{(L{+}\frac 1 2)} \left(u,\frac{v(s+1)}{v s+1},\frac{s+w}{s+1}\right)
\end{split}
\end{equation}
We remark that the actual integrals with even weight are symmetric in exchange of $u$ and $v$ (the integral has a symmetry axis); we introduce odd-weight objects which break the symmetry, but we can alternatively write down recursion with $u, v$ swapped, which give different odd-weight functions but will not affect the even-weight integrals. Nicely, the recursion applies to $L=0$, where the tree case is defined to be $I_{\rm pb}^{(0)}=1-u-v+u v w$. By applying the first equation of ~\eqref{penta} to this tree result, we obtain a weight-$1$ function $I_{\rm dp}^{(1/2)}=\frac{1-u-v+ u v w}{1- uw}\log(u w)$, and by applying the second equation, we arrive at the well-known one-loop chiral-pentagon $I_{\rm pb}^{(L=1)}:=I_{\rm p}(u,v,w)$:
\begin{equation*}
I_{\rm p}(u,v,w)
=\log u\log v+\Li_2(1{-}u)+\Li_2(1{-}v)+\Li_2(1{-}w)-\Li_2(1{-}uw)-\Li_2(1{-}v w).
\end{equation*}
The recursion makes it manifest that the result will always be pure functions starting $L=1$. We emphasize that by using the algorithm of~\cite{CaronHuot:2011kk}, it is straightforward to compute the symbol of $I_{\rm pb}^{(L)}$ to any loop order.

We are mainly interested in the alphabet of the resulting symbol. As we have seen at $L=1$ (weight $2$), and in fact also for $L=\frac{3}{2}$ (weight $3$) as obtained using the first line of \eqref{penta}, the alphabet consists of eight letters, $u,v, w, 1-u, 1-v, 1-w, 1-u w, 1-v w$. However, these are just degenerate cases, and starting at $L=2$ we find $9$ letters where the additional one is nothing but the tree-level factor $1-u-v+ u v w$:
\begin{equation}\label{pbalp}
{\bf A} [I_{\rm pb}^{(L)}]=\{u, v, w, 1-u, 1-v, 1-w, 1-u w, 1-v w, 1-u-v+u v w \}.
\end{equation}
We have checked up to $L=5$ (weight $10$), and in sec. \ref{sec3} we will give an all-order proof using the recursion that this $9$-letter alphabet is true to all loops with $L\geq 2$.  

Now we identify the alphabet ${\bf A}[I_{\rm pb}^{(L)}]$ with that of $D_3\simeq A_3$ cluster algebra, and we do so in two ways. First, similar to \cite{Chicherin:2020umh}, we find the bi-rational change of variables
\begin{equation}\label{change}
u=\frac 1 {1+z_2}\,,\quad v=\frac 1 {1+z_3}\,, \quad w=1+z_1
\end{equation}
and, up to multiplicative redefinition, the alphabet \eqref{pbalp} becomes
\begin{equation}
{\bf A}[I_{\rm pb}^{(L)}] \simeq \{z_1, z_2, z_3, 1+z_1, 1+z_2, 1+z_3, z_1-z_2, z_1-z_3, z_1+z_2 z_3\}.
\end{equation}
which we immediately recognize as that of $D_3$ (second line of \eqref{AD} with $d=3$). A trivial change of variables turns it into that of $A_3$ (first line of \eqref{AD} with $d=3$).

These changes of variables may seem a bit arbitrary, but as mentioned earlier we can reach at the conclusion in a totally invariant way. Pick any {\it positive} parametrization of the $9$ letters in \eqref{pbalp}; 
it does not matter what positive parametrization we choose as long as they give subtraction-free polynomials $p_I$ for $I=1,\dots,9$, and we write the stringy canonical form
\begin{equation}
I_{\{p\}}(\{s\})=(\alpha')^3 \int_{\mathbb{R}_{>0}^3} \prod_{i=1}^3 d\log x_i \prod_{I=1}^9 p_I(x_1,x_2, x_3)^{\alpha' s_I}\,.
\end{equation}
Without being smart, we can simply compute the Minkowski sum of Newton polytopes of $p_I$'s which gives the convergence domain of $I_{\{p\}}$, and we find that the result is nothing but a $3$-dimensional associahedron! It is given by $9$ inequalities of the form $X_a (\{s\})\geq 0$, each of which can be written as a linear combination of the $s_I$'s. With these $9$ linear combinations, we can identify the $9$ ${\bf u}$ variables of $A_3$ by writing $p_I^{\alpha^\prime s_I}=\prod_{(i,j)} {\bf u}_{i,j}^{\alpha^\prime X_{i,j}}$. With a bit hindsight, we label the ${\bf u}$'s by diagonals of a hexagon as in \eqref{uAn}, and they automatically satisfy the $9$ ${\bf u}$ equations in \eqref{ueqAn}. This gives a description of the $A_3$ alphabet that is totally parametrization-independent, ${\bf A}[I_{\rm pb}^{(L)}]=\{{\bf u}_{i,j}| 1\leq i<j-1<6,~(i,j)\neq (1,6)\}$, where the ${\bf u}$ variables are multiplicative combinations of the original letters:
\begin{align}\label{uexp}
&{\bf u}_{1,3}=w,\, {\bf u}_{1,4}=\frac{1-v}{1-v w},\, {\bf u}_{1,5}=\frac{u(1-v w)}{1-v}, \, {\bf u}_{2,4}=1-v w,\,{\bf u}_{2,5}=\frac{1-w}{(1-u w)(1-v w)},\nonumber\\
& {\bf u}_{2,6}=1-u w,\, {\bf u}_{3,5}=\frac{v(1-u w)}{1-u},\, {\bf u}_{3,6}= \frac{1-u}{1-u w},\, {\bf u}_{4,6}=\frac{1-u-v+u v w}{(1-u)(1-v)}
\end{align}
One can easily check that \eqref{uexp} satisfies ${\bf u}$ equations and with any positive parametrization all ${\bf u}$ variables are between $0$ and $1$. 

As mentioned such a description using ${\bf u}$ variables is very useful, {\it e.g.} each of the $9$ boundaries can be obtained by sending exactly one ${\bf u}$ to zero. There are $6$ boundaries by ${\bf u}_{i, i{+}2} \to 0$ (for $i=1, \cdots, 6$) which are $A_2$ (pentagon), and $3$ boundaries by ${\bf u}_{i, i{+}3} \to 0$ (for $i=1,2,3$) which are $D_2\simeq A_1^2$ (quadrilateral). However, for $I_{\rm pb}^{(L)}$ many boundaries are too degenerate as the symbol vanishes identically, only the following $4$ boundaries correspond to non-trivial limits: ${\bf u}_{1,3}\to 0$, ${\bf u}_{1,5}\to 0$, ${\bf u}_{3,5}\to 0$, which are $A_2$, and ${\bf u}_{2,5}\to 0$, which is $A_1$. 

The first $A_2$ is given by $w\to 0$, and it is the familiar collinear limit which gives seven-point penta-box ladder with alphabet $\{u, v, 1-u, 1-v, 1-u-v\}$~\cite{Dixon:2020cnr}; the next two $A_2$ are also collinear limits reached by $u\to 0$ and $v \to 0$ respectively, though the integral diverges in these limits. Finally, the last limit is given by $w\to 1$ which can be nicely reached by reducing the kinematics to two dimensions! In such a limit, we have $u\to u^{\rm 2d}_{2,4,6,8}$ and $v\to u^{\rm 2d}_{1,3,5,7}$, and it turns out that the resulting $D_2$ function is even simpler than box ladder \eqref{fL}. As first noted in~\cite{He:2020uxy} from resummation, penta-box ladder in 2d is perhaps the simplest $A_1^2$ function: it is given by the product of weight-$L$ classical polylogarithm function of $u$ and that of $v$,
\[
I_{\rm pb}^{(L)} \to F(u) F(v)\,,\quad F(u):=\Li_L(1-u^{-1})
\]
{\it e.g.} for $L=1$ we have the chiral pentagon in 2d: $I_{\rm p}=\log(u)\log (v)$. 
\subsection{$D_4$ for double-penta-ladder integrals}

Finally, we move to $I^{(L)}_{\rm dp}(x_1, x_2, x_4, x_5, x_6, x_7)$, which depends on $4$ cross-ratios~\footnote{We use $v_i$ for $i=1,2,3,4$, which was referred to as $u_i$ for $i=1,2,3,4$ in~\cite{He:2020uxy}, to avoid confusion with ${\bf u}$ variables of cluster algebra.}
\begin{equation}\label{v1v2v3v4}
v_1=\frac{x_{16}^2 x_{25}^2}{x_{15}^2 x_{26}^2}
,\ 
v_2=\frac{x_{14}^2 x_{57}^2}{x_{15}^2 x_{47}^2}
,\ 
v_3=\frac{x_{27}^2 x_{46}^2}{x_{26}^2 x_{47}^2}
,\  
v_4=\frac{x_{15}^2 x_{24}^2}{x_{14}^2 x_{25}^2}
.
\end{equation}
As shown in~\cite{He:2020uxy} for $L\geq 1$ it satisfies a similar recursion~\footnote{Again the integrals have $v_1 \leftrightarrow v_2$ symmetry, but not the odd-weight functions; we could use the recursion with $v_1 \leftrightarrow v_2$ which does not affect the even-weight integrals.}:
\begin{equation}\label{double1}
\begin{split}
I_{\rm dp}^{(L+\frac12)}(v_1,v_2,v_3,v_4)&=\int_0^\infty {\rm d}\log\frac{t{+}1}{t} I_{\rm dp}^{(L)}\biggl(\frac{v_1(t{+}v_4)}{t{+}v_1 v_4},v_2,\frac{t v_3}{t{+}v_1 v_4},\frac{v_4(t{+}1)}{t{+}v_4}\biggr),\\
I_{\rm dp}^{(L{+}1)}(v_1,v_2,v_3,v_4)&=\int_0^\infty {\rm d}\log(s{+}1)I_{\rm dp}^{(L+\frac12)} \biggl(v_1,\frac{v_2(s{+}1)}{v_2 s{+}1},\frac{v_3}{1{+}s v_2},\frac{s{+}v_4}{s{+}1}\biggr),
\end{split}
\end{equation}
where the one-loop case $I_{\rm dp}^{(1)}$, is a seven-point chiral hexagon evaluating to
\begin{equation*}
        \begin{split}
     \begin{tikzpicture}[baseline={([yshift=-.5ex]current bounding box.center)},scale=0.18]
                \draw[black,thick] (0,0)--(4,0)--(6,3.46)--(4,6.93)--(0,6.93)--(-2,3.46)--cycle;
                \draw[black,thick] (0,6.93)--(-1,8.66);
                \draw[black,thick] (4,6.93)--(5,8.66);
                \draw[black,thick] (7.74,2.46)--(6,3.46)--(7.74,4.46);
                \draw[black,thick] (4,0)--(5,-1.73);
                \draw[black,thick] (0,0)--(-1,-1.73);
                \draw[black,thick] (-2,3.46)--(-4,3.46);
                \draw[decorate, decoration=snake, segment length=12pt, segment amplitude=2pt, black,thick] (0,7)--(0,0);
                \draw[decorate, decoration=snake, segment length=12pt, segment amplitude=2pt, black,thick] (4,0)--(4,6.93);
                \filldraw[black] (-1,8.66) node[anchor=south east] {{7}};
                \filldraw[black] (5,8.66) node[anchor=south west] {{1}};
                \filldraw[black] (7.74,4.46) node[anchor=west] {{2}};
                \filldraw[black] (7.74,2.46) node[anchor=west] {{3}};
                \filldraw[black] (5,-1.73) node[anchor=north west] {{4}};
                \filldraw[black] (0,-1.73) node[anchor=north east] {{5}};
                \filldraw[black] (-4,3.46) node[anchor=east] {{6}};
            \end{tikzpicture}&= \log v_1\log v_2-\Li_2(1)+\Li_2(1-v_1)+\Li_2(1-v_2) \\[-4ex]
            &+ \Li_2(1-v_4)-\Li_2(1-v_1 v_4)-\Li_2(1-v_2 v_4)+\Li_2(1-v_3),
        \end{split}\label{hexagon}
    \end{equation*}
Note that as also noticed in~\cite{He:2020uxy}, although we can similarly find $L=\frac 1 2$ and $0$ cases, the latter (tree case) will not be simply weight-$0$ object but also involves $\log$ term. 

We have computed up to $L=4$ and find the alphabet of penta-box-ladder integrals as (for $L\geq 2$)
\begin{align}\label{D4}
{\bf A}[I_{\rm dp}^{(L)}]=\biggl\{& v_1, v_2, v_3, v_4, 1-v_1, 1-v_2, 1-v_3, 1-v_4, 1-v_1 v_4, 1-v_2 v_4, \\\nonumber
&1-v_3-v_1 v_4, 1-v_3-v_2 v_4, \frac{1-x_+}{1-x_-}, \frac{v_4-x_+}{v_4-x_-}, \frac{v_1^{-1}-x_+}{v_1^{-1}-x_-}, \frac{v_2^{-1}-x_+}{v_2^{-1}-x_-}\biggr\}
\end{align}
where we have defined $x_{\pm}:=\frac{-1+v_1+v_2+v_3+v _1 v_2 v_4\pm\sqrt{\Delta_7}}{2 v_1 v_2}$ with the seven-point Gram determinant  $\Delta_7=({-}1{+}v_1{+}v_2{+}v_3{+}v_1 v_2 v_4)^2{-}4 v_1 v_2 v_3 (1{-}v_4)$.~\footnote{In v1 of ~\cite{He:2020uxy}, we also included $\frac{1-v_1-v_2+v_1 v_2 x_+}{1-v_1-v_2+ v_1 v_2 x_-}$ which is not multiplicatively independent.} Only the first $10$ letters in \eqref{hexagon} appear for $L=1$, and as can be obtained from \eqref{double1}, an additional letter $1-v_3- v_2 v_4$ appears at $L=\frac{3}{2}$ (weight $3$); these are all degenerate cases of the alphabet \eqref{D4}, which become generic starting $L=2$. 

To identify the alphabet with that of $D_4$, it is crucial to have any change of variables which gets rid of the square root. Clearly this can be done using momentum twistors ({\it c.f. } \cite{DelDuca:2011wh}). Here we adopt the following parametrization of the $7$ momentum twistors involved:
\begin{align}\label{Z7s}
{\bf Z}=\left(
\begin{array}{ccccccc}
 0 & 1 & 1 & 0 & 1 & 0 & 1 \\
 1 & 0 & 1 & 0 & 0 & 0 & a_4 \\
 0 & a_1 & 0 & 1 & 1 & 0 & 0 \\
 0 & a_2 & 0 & 0 & a_3 & 1 & 1 \\
\end{array}
\right)
\end{align}
By plugging \eqref{Z7s} into \eqref{v1v2v3v4} we find 
\begin{equation*}
v_1=\frac{a_2-a_3}{(1-a_1) (1-a_3)}\,,\,
v_2=\frac{-a_4}{(1-a_3)(1-a_4)}\,, \,
v_3=\frac{-a_1}{(1-a_1)(1-a_4)}\,, \,
v_4=\frac{a_2(1-a_3)}{(a_2-a_3)},
\end{equation*}
and $\Delta_7$ above becomes a perfect square. The upshot is that the alphabet becomes multiplicative combinations of exactly $16$ polynomials:
\begin{align}\label{dpletterina}
    {\bf A}[I_{\rm dp}^{(L)}]= \{& 1-a_1, a_1, 1-a_2, a_2, 1-a_3,a_3, 1-a_4, a_4, 1-a_1-a_2, a_2-a_3, \nonumber\\
    & 1-a_1-a_2+a_1 a_3, 1-a_4+a_1 a_4, 1-a_3+a_3 a_4,\\
    & 1-a_2-a_4+a_1 a_4+a_2 a_4, a_2-a_3+a_3 a_4, a_2-a_3-a_1 a_3 a_4+a_3 a_4\}\nonumber
\end{align}
From here it is straightforward to find a {\it positive parametrization}: we simply need to find positive variables which guarantees all $a_i<0$ for $i=1,2,3,4$ as well as $a_2>a_3$, {\it e.g.} $a_1=-x_1, a_2=-x_2, a_3=-x_2-x_3, a_4=-x_4$ and the above $16$ polynomials become {\it subtraction-free}. Then we can simply follow the procedure by computing the Minkowski sum of the Newton polytopes of such $16$ polynomials, and remarkably we find a $D_4$ polytope! It is then straightforward to work out the $16$ ${\bf u}$ variables as an invariant description of ${\bf A}[I_{\rm dp}^{(L)}]$, \eqref{uDn} for $n=4$, provided that they satisfy the following $4+4+8=16$ ${\bf u}$ equations:
\begin{align}\label{ueqD4}
& 1-{\bf u}_{1,2}={\bf u}_3 {\bf \tilde u}_3 {\bf u}_4 {\bf \tilde u}_4 {\bf u}_{3,4}^2 {\bf u}_{2,3} {\bf u}_{2,4} {\bf u}_{4,1} {\bf u}_{3,1}\,,\quad \&~{\rm cyclic}\nonumber\\
&1-{\bf u}_{1,3}={\bf u}_4 {\bf \tilde u}_4 {\bf u}_{4,1} {\bf u}_{4,2} {\bf u}_{2,4} {\bf u}_{3,4}\,, \quad \&~{\rm cyclic}\nonumber\\
&1-{\bf u}_1={\bf \tilde u}_2 {\bf \tilde u}_3 {\bf \tilde u}_4 {\bf u}_{2,3} {\bf u}_{3,4} {\bf u}_{2,4}\,\quad \&~{\rm cyclic}~\&~({\bf u} \leftrightarrow {\bf \tilde u}). 
\end{align}
where on the last line we have ${\bf u} \leftrightarrow {\bf \tilde u}$ as well. Explicit expressions of ${\bf u}$ variables in terms of $v_1, \cdots, v_4$ involve square roots but they simplify in terms of $a_1, \cdots, a_4$:

\vspace{-2.5ex}
{\footnotesize
\begin{align}\label{uD4}
&{\bf u}_1=\frac{a_1 (a_3{-}1)}{a_3 a_1{-}a_1{-}a_2{+}1},{\bf u}_2= \frac{a_1{+}a_2{-}1}{a_1{-}1},{\bf u}_3= \frac{a_3 (a_1 a_4{-}a_4{+}1)}{{-}a_2{+}a_3{+}a_1 a_3 a_4{-}a_3 a_4},{\bf u}_4= \frac{{-}a_2{+}a_3{-}a_3 a_4}{(a_2{-}a_3) (a_4{-}1)},\nonumber\\
&{\bf \tilde u}_1= \frac{a_2{-}a_2 a_3}{a_2{-}a_3},{\bf \tilde u}_2= \frac{{-}a_1{-}a_2{+}1}{a_3 a_1{-}a_1{-}a_2{+}1},{\bf \tilde u}_3= \frac{a_1 a_4{-}a_4{+}1}{(a_1{-}1) (a_4{-}1)},{\bf \tilde u}_4= \frac{a_2{-}a_3{+}a_3 a_4}{a_2{-}a_3{-}a_1 a_3 a_4{+}a_3 a_4},\nonumber\\
&{\bf u}_{13}= \frac{(a_1{+}a_2{-}1) a_4}{a_4 a_2{-}a_2{+}a_1 a_4{-}a_4{+}1},{\bf u}_{24}= {-}\frac{(a_2{-}1) (a_1 a_4{-}a_4{+}1)}{a_4 a_2{-}a_2{+}a_1 a_4{-}a_4{+}1},{\bf u}_{31}= \frac{a_2{-}a_3{+}a_3 a_4}{a_4 a_3{-}a_3{+}1},{\bf u}_{42}= \frac{1{-}a_3}{a_4 a_3{-}a_3{+}1},\nonumber\\
&{\bf u}_{12}= \frac{a_3 a_1{-}a_1{-}a_2{+}1}{(a_1{+}a_2{-}1) (a_3{-}1)},{\bf u}_{23}= \frac{(a_1{-}1) (a_4 a_2{-}a_2{+}a_1 a_4{-}a_4{+}1)}{(a_1{+}a_2{-}1) (a_1 a_4{-}a_4{+}1)},\nonumber\\
&{\bf u}_{34}= \frac{(a_4{-}1) ({-}a_2{+}a_3{+}a_1 a_3 a_4{-}a_3 a_4)}{(a_1 a_4{-}a_4{+}1) (a_2{-}a_3{+}a_3 a_4)},{\bf u}_{41}= {-}\frac{(a_2{-}a_3) (a_4 a_3{-}a_3{+}1)}{(a_3{-}1) (a_2{-}a_3{+}a_3 a_4)}\,,
\end{align}}
and we can easily check that they satisfy the $16$ equations \eqref{ueqD4}.

From \eqref{uD4} it is also straightforward to rewrite the ${\bf u}$ variables in terms of Pl{\"u}cker coordinates of $G(4,7)$:

\vspace{-2.5ex}
{\footnotesize
\begin{align*}
\biggl\{&\tilde {\bf u}_1= \frac{\langle 1234\rangle \langle 1457\rangle}{\langle 1245\rangle \langle 1347\rangle},
\tilde {\bf u}_2= -\frac{\langle 1(27)(34)(56)\rangle}{\langle 1257\rangle \langle 1346\rangle},
\tilde {\bf u}_3= -\frac{\langle 6(12)(34)(57)\rangle}{\langle 1256\rangle \langle 3467\rangle},\tilde {\bf u}_4= -\frac{\langle 1567\rangle \langle 4(12)(35)(67)\rangle}{\langle 1467\rangle \langle 5(12)(34)(67)\rangle},\\
&{\bf u}_1= \frac{\langle 1267\rangle \langle 1457\rangle}{\langle 1257\rangle \langle 1467)},
{\bf u}_2= -\frac{\langle 1(27)(34)(56)\rangle}{\langle 1256\rangle \langle 1347\rangle},
{\bf u}_3= -\frac{\langle 1345\rangle \langle 6(12)(34)(57)\rangle}{\langle 1346\rangle \langle 5(12)(34)(67)\rangle},{\bf u}_4= -\frac{\langle 4(12)(35)(67)\rangle}{\langle 1245\rangle \langle 3467\rangle},\\
&{\bf u}_{12}= -\frac{\langle 1257\rangle \langle 1347\rangle \langle 1456\rangle}{\langle 1457\rangle \langle 1(27)(34)(56)\rangle},{\bf u}_{23}= \frac{\langle 1256\rangle \langle 1346\rangle \langle 7(12)(34)(56)\rangle}{\langle 1(27)(34)(56)\rangle \langle 6(12)(34)(57)\rangle},\\
&{\bf u}_{34}= \frac{\langle 1246\rangle \langle 3467\rangle \langle 5(12)(34)(67)\rangle}{\langle 4(12)(35)(67)\rangle \langle 6(12)(34)(57)\rangle},{\bf u}_{41}= -\frac{\langle 1245\rangle \langle 1467\rangle \langle 3457)}{\langle 1457\rangle \langle 4(12)(35)(67)\rangle},{\bf u}_{42}= \frac{\langle 1457\rangle \langle 3456\rangle}{\langle 1456\rangle \langle 3457\rangle},\\
&{\bf u}_{13}= -\frac{\langle 4567\rangle \langle 1(27)(34)(56)\rangle}{\langle 1456\rangle \langle 7(12)(34)(56)\rangle},{\bf u}_{24}= -\frac{\langle 1247\rangle \langle 6(12)(34)(57)\rangle}{\langle 1246\rangle \langle 7(12)(34)(56)\rangle},
{\bf u}_{31}= -\frac{\langle 4(12)(35)(67)\rangle}{\langle 1246\rangle \langle 3457\rangle}
\biggr\}
\end{align*}
}

Not surprisingly, this amounts to embed the $D_4$ alphabet naturally in the $E_6 \sim G(4,7)/T$, and we remark that the embedding is not unique. In fact, we can first write the $42$ ${\bf u}$ variables for $E_6$ cluster algebra, in terms of Pl{\"u}cker coordinates in $G(4,7)$. Then we express our $16$ letters above as monomials of the $42$ variables, and we find there are four solutions, each corresponding to a co-dimension $2$ boundaries of the $E_6$ space. What we have found above is just one of the four solutions and the other three can be obtained by cyclic rotation in $D_4$. 

Let us also identify all $16$ boundaries of the $D_4$ alphabet, each reached by one ${\bf u}\to 0$. There are $12$ $A_3$ corresponding to the first $12$ ${\bf u}$ variables of \eqref{ueqD4} and $4$ $A_1^3$ for the last $4$. For $I_{\rm dp}^{(L)}$, we find that the symbol vanishes for $4$ boundaries, namely those for ${\bf u}_2, {\bf u}_4, {\bf u}_{12}, {\bf u}_{41}$, which are too degenerate. We believe that all the remaining $12$ non-trivial boundaries have certain physical interpretation. Note $I_{\rm dp}^{(L)}$ diverges at those for ${\bf u}_{23}$,${\bf u}_{34}$ ($A_1^3$) and those for ${\bf \tilde u}_2$, ${\bf \tilde u}_3$, ${\bf \tilde u}_4$, ${\bf u}_{13}$, ${\bf u}_{31}$ ($A_3$), thus it remains finite at $5$ boundaries (all $A_3$): ${\bf u}_1 \to 0$, ${\bf \tilde u}_1 \to 0$, ${\bf u}_3 \to 0$, ${\bf u}_{24} \to 0$ and ${\bf u}_{42} \to 0$. 

At least the physical interpretation of the first two boundaries are very clear: the first one corresponds to $v_3 \to 0$, and the integral reduces to $I_{\rm pb}^{(L)}$ with with $(u,v,w)=(u_1, u_2, u_4)$, while the second one correspond to $v_4\to 0$ and it reduces to the six-point double-penta ladder~\cite{Drummond:2010cz}, which is the famous $A_3$ for hexagon function\cite{Dixon:2011pw}. As one can see from \eqref{ueqD4}, the first $D_3 \simeq A_3$ sub-algebra has $9$ remaining ${\bf u}$ variables, and in the order of \eqref{uexp} they read ${\bf \tilde u}_1$, ${\bf u}_{41}$, ${\bf u}_{31}$, ${\bf u}_4$,${\bf u}_3$, ${\bf u}_2$,${\bf u}_{13}$, ${\bf u}_{12}$, ${\bf u}_{42}$. The $9$ remaining variables for the second $A_3$ are obtained by swapping ${\bf u}$ and ${\bf \tilde u}$, which are combinations of the familiar $9$ letters for hexagon bootstrap (in terms of original variables, they are $v_1, v_2, v_3, 1-v_1, 1-v_2, 1-v_3$ and three involving $x_{\pm}$ with $v_4 \to 0$). 

Last but not least, we note that our variables $a_1, a_2, a_3, a_4$ are just relabelling of $y_2, y_3, y_4, y_5$ in \cite{Chicherin:2020umh}, thus alternatively one can find the following bi-rational transformation which sends them to the familiar $z$ variables for $D_4$:
\begin{equation}
a_1=\frac{z_1 (1+z_4)}{z_1-z_4}\,,\quad a_2=\frac{z_4 (z_1-z_2)}{z_2 (z_1-z_4)}\,,\quad a_3=\frac{z_4}{z_2}\,,\quad a_4=-\frac{(z_2-z_4)}{z_4(1+z_3)}\,.\nonumber
\end{equation}
and the alphabet becomes the $16$ polynomials on the second line of \eqref{AD} with $d=4$.

\subsection{Comments on ``adjacency" constraints on cluster function spaces}

Let us describe constraints on adjacent entries that we observe for the symbol of the ladder integrals, which may have their origins in extended Steinmann relations~\cite{Caron-Huot:2016owq, Dixon:2016nkn, Caron-Huot:2019bsq} or cluster adjacency~\cite{Drummond:2017ssj, Drummond:2018caf}, and briefly comment on their consequences on cluster function spaces. 

To describe these constraints, it is important to use ``good" variables: it turns out that for both $I_{\rm pb}^{(L)}$ and $I_{\rm pb}^{(L)}$, the function/symbol expressed in the $z$ variables is much shorter than that in terms of cross-ratios (or ${\bf u}$ variables).  We also note that the $z$ variables also exactly give $3$ and $4$ combinations that appear in the last entry of these integrals (this point will be important when considering differential equations or resummation). Therefore, we consider such constraints using $z$ variables, first for $I_{\rm pb}^{(L)}$. As we have checked through $L=5$, $12$ pairs never appear next to each other in its symbol: there is no $a\otimes a$ for $a=1+z_1, 1+z_2, 1+z_3$ or $z_1+ z_2 z_3$, and there is no $a \otimes b$ or $b\otimes a$ for $\{a,b\}$ equals 
\begin{align}\label{A3constraint}
   &\{z_1, 1{+}z_2\}, \{z_1, 1{+}z_3\}, \{z_2, z_1{-}z_3\}, \{z_3, z_1{-}z_2\}, \{z_1{-}z_2, z_1{-}z_3\},\nonumber\\
   &\{z_2, z_3\}, \{z_2, 1{+}z_3\}, \{1{+}z_2, z_3\}\,. 
\end{align}
Note that these constraints imply that for the original $16$ letters, none of $\{w, w\}$, $\{1-u-v+u v w, 1-u-v +u v w\}$, $\{1-u, 1-v\}, \{1-u, 1-v w\}, \{1-v, 1-u w\}$ or $\{1-u w, 1-v w\}$ is allowed, but these are much weaker than those in $z$ var.

One can construct the corresponding cluster function space at symbol level to relatively high weight, and we focus on the $A_3$ space with physical first-entry condition, {\it i.e.} the collection of all weight-$k$ integrable symbols with only $u,v,w$ in the first entry. The first observation is that the dimension of the space (denoted as $d_k$) is $3, 11, 40, 146, 538, 2006, \cdots$, and we conjecture in general it reads 
\[d_k=\frac{3}{2}\times 4^{k-1}+2\times 3^{k-1}-2^{k-2},\]
which is a special case of a more general observation on the dimension of such spaces. Now if we impose the adjacency conditions in terms of $z$ variables, we find that the dimension of the space is drastically reduced to $3,8, 20, 44, 88, 171,\cdots$ (note that for weight $6$ the space is reduced by more than $90$ percent!). At least up to $L=3$, it should be easy to bootstrap penta-ladder integrals by imposing other conditions.

Similarly, for $I_{\rm pb}^{(L)}$ through $L=4$, we find that there are $37$ pairs that cannot appear next to each other in the symbol (some of them may become allowed at higher loops). First $z_1-z_2-z_1 z_2+ z_1 (z_3+z_4)-z_3 z_4$ cannot appear next to itself, $z_2$, $1+z_2$, $z_3$, $z_4$ or $z_1 + z_3 z_4$; $z_1+z_3 z_4$ also cannot appear next to $1+z_1$, $z_2$, $1+z_2$, $z_2-z_3$ or $z_2-z_4$; $z_2+z_3 z_4$ cannot appear next to $1+z_1$, $1+z_2$, $z_1-z_3$ or $z_1-z_4$; the remaining pairs that are not allowed read

\vspace{-2ex}
{\footnotesize \begin{align*} &\left\{z_1+1,z_2\right\},\left\{z_1+1,z_3\right\},\left\{z_1+1,z_2-z_3\right\},\left\{z_1+1,z_4\right\},\left\{z_1+1,z_2-z_4\right\},\left\{z_2,z_1-z_3\right\},\\
&\left\{z_2+1,z_3\right\},\left\{z_2+1,z_1-z_3\right\},\left\{z_2+1,z_4\right\},\left\{z_2+1,z_1-z_4\right\},\left\{z_1-z_2,z_1-z_2\right\},\left\{z_3,z_1-z_4\right\},\\
&\left\{z_3,z_2-z_4\right\},\left\{z_3+1,z_3+1\right\},\left\{z_1-z_3,z_4\right\},\left\{z_1-z_3,z_1-z_4\right\},\left\{z_1-z_3,z_2-z_4\right\},\left\{z_2-z_3,z_4\right\},\\
&\left\{z_2-z_3,z_1-z_4\right\},\left\{z_2-z_3,z_2-z_4\right\},\left\{z_4+1,z_4+1\right\},\left\{z_2,z_1-z_4\right\}
\end{align*}}%
We have not attempted to find physical explanation for these adjacency conditions, which may have origin from extended-Steinmann relations; they could be alternatively explained by re-expressing our variables in terms of the $E_6$ variables suitable for cluster adjacency~\cite{Drummond:2017ssj, Drummond:2018caf}. We leave a more systematic study of these conditions and their consequences for cluster function spaces to a future work.

\section{Cluster-algebra alphabets from recursive ${\rm d}\log$ forms}\label{sec3}

Having identified alphabets of three classes of ladder integrals with $D_2, D_3, D_4$, a natural question is why these alphabets stay invariant as $L$ increases? We do not have a complete answer, but here we would like to sketch an argument based on the recursive $d\log$ forms, which also allows us to further extend our explorations.

Let us illustrate the idea by the following example. Suppose one of the terms in the symbol integration reads
\begin{equation}
 \int {\rm d}\log(t{+}b_1) (t{+}b_2) \otimes a \otimes (t{+}b_3) \otimes (t{+}b_4)+\cdots.
\end{equation}
Here $a,b_1,b_2,b_3,b_4$ are different constants, $t$ is integrated over the region $\mathbb{R}_+$. According to the algorithm of symbol integration, generically the result depend on letters of the form:
\[\{a;\ b_1,\ b_2,\ b_3,\ b_4;\ b_1{-}b_2,\ b_1{-}b_3,\ b_1{-}b_4,\ b_2{-}b_3,\ b_2{-}b_4,\ b_3{-}b_4\}.
\]
Note that constant $a$, which shows up in the alphabet, does not mix with any $b_i$'s, and in addition to $b_i$ from each $t+b_i$, each pair of factors $t+b_i$ and $t+b_j$ contributes a $b_i-b_j$ to the final alphabet. This follows directly from our algorithm reviewed earlier. Each $t+b_i$ contributes $b_i$ to the alphabet of the result when evaluated at the end point, following the first part of the algorithm. A constant $a$ is produced both from end-point value and from the situation of $(F(t)\otimes a){\rm d}\log(t+b_i)$. Finally, whenever the last entry reads $t+b_i$ with ${\rm d}\log(t+b_1)$ as differential form, it contributes $b_i-b_1$ as the new last entry, with the differential form changed to ${\rm d}\log\frac{t-b_i}{t-b_1}$. Since $t+b_i$ becomes a new entry of ${\rm d}\log$ form, recursively all the mixing letters $b_i-b_j$ should appear in the final alphabet.

We can make an {\it estimation} for the alphabet of any integral of this type without performing the symbol integration explicitly. If we have an $t$-deformed alphabet which also include possible factors of ${\rm d}\log(t+b)$ (for our purpose the latter only involves $b=0,1$)
\[
\{a_1,\cdots\ a_n;\ t+b_1,\cdots, t+b_m\}
\]
Then after such integration, all possible letters must be in the collection
\begin{equation}\label{generalletter}
\{a_i\}_{i=1,\cdots,n}\cup\{b_i\}_{i=1,\cdots,m}\cup\{b_i-b_j\}_{i,j=1,\cdots, m} 
\end{equation}
We should emphasize that some of the mixing letters can be {\it spurious}, since any two letters $t+b$ and $t+b'$ will not contribute $b-b'$ unless they show up in the {\it same term} of the integrand. Therefore \eqref{generalletter} is just an {\it estimation}, which gives an upper bound of the actual alphabet. 

With this simple estimation, we can already show that the alphabet of penta-box ladder must be $D_3$. By applying \eqref{generalletter} recursively, we show that the upper bound is \eqref{pbalp} to all loops. In the first step of the recursion, after the deformation \eqref{penta}, the $9$ letters become
\[
\left\{\frac{u(t{+}w)}{t{+}u w},v,\frac{w(t{+}1)}{t{+}w},\frac{t(1{-}u)}{t{+}u w},1{-}v,\frac{t(1{-}w)}{t{+}w},\frac{t(1{-}u w)}{t{+}u w},\frac{(1{-}v w)t{+}w(1{-}v)}{t{+}w},\frac{t(1{-}u{-}v{+}u v w)}{t{+}u w}\right\}
\]
which after expansion of the symbol gives the following alphabet
\[
\left\{u,v,w,1{-}u,1{-}v,1{-}w,1{-}u w,1{-}v w,1{-}u{-}v{+}u v w;\ t,t{+}1,t{+}w,t{+}u w,t{+}\frac{w(1{-}v)}{1{-}v w}\right\}.
\]
So all the $a_i$ type letters are just the original ones. Since ${\rm d}\log$ form we are dealing with is just ${\rm d}\log\frac{t{+}1}{t}$, there are no new $t+b_i$ type letters we need to add in the alphabet above. All the letters linear in $t$ will then produce letters of $b_i$ type as $\left\{0,1,w,u w,\frac{w(1{-}v)}{1{-}v w}\right\}$ and $b_i-b_j$ type as
\begin{equation*}
\begin{split}
\biggm\{1,w,u w,\frac{w(1{-}v)}{1{-}v w};1{-}w,1{-}u w,1{-}\frac{w(1{-}v)}{1{-}v w}\left(=\frac{1{-}w}{1{-}v w}\right); w{-}u w(=w(1{-}u)),\\
w-\frac{w(1{-}v)}{1{-}v w}\left(=\frac{v w(1{-}w)}{1{-}v w}\right); u w-\frac{w(1{-}v)}{1{-}v w}\left(=-\frac{w(1{-}u{-}v{+}u v w)}{1{-}v w}\right) \biggm\}
\end{split}
\end{equation*}
which contribute no new factors to $I_{\rm pb}^{(L{+}\frac 1 2)}$, besides the original $9$ letters of $I_{\rm pb}^{(L)}$. Similarly we find the alphabet also stays invariant from $I^{(L{+}\frac12)}_{\rm pb}$ to $I^{(L{+}1)}_{\rm pb}$. We have also checked up to weight $10$ that the actual alphabet is \eqref{pbalp}. 

We can also use this estimation for $D_4$ cases {\it i.e.} double-penta-ladders $I^{(L)}_{\rm dp}$. From $I^{(L)}_{\rm dp}$ to $I^{(L{+}\frac12)}_{\rm dp}$, this argument shows that the $16$ letters \eqref{dpletterina} produce $20$ letters from the recursion. Besides $16$ original letters, the new ones read
\begin{equation*}
\begin{split}
\{1{-}a_2{+}a_1a_3a_4,\ 1{-}a_2{-}a_4{+}a_1a_4{+}a_2a_4{-}a_1a_3a_4,\ 1{-}a_1{-}a_2{+}a_1a_3{-}a_1a_3a_4,\\
1{-}a_2{-}a_4{+}a_1a_4{+}a_2a_4{-}a_1a_3a_4{+}a_1a_3a_4^2\}.
\end{split}
\end{equation*}
However, as mentioned we have checked that through weight $8$ the alphabet of $I^{(L)}_{dp}$ stays invariant, so that the new ones must be spurious at least up to $L\leq4$. Take $I^{(2)}_{dp}$ as an example: the last two letters there are trivially spurious, since the corresponding $t+b_i$ and $t+b_j$ that produce them never appear in the same term; we also find the first two letters got cancelled in the final result. Similarly, if we start again with $16$ letters of $I^{(L{+}\frac12)}_{\rm dp}$, the recursion for $I^{(L{+}1)}_{\rm dp}$ also produces four new letters
\begin{equation*}
\begin{split}
\{1{-}a_2{+}a_1a_3a_4,\ 1{-}a_1{-}a_2{+}a_1a_3{-}a_1a_3a_4,\ 1{-}a_3{+}a_3a_4{-}a_1a_3a_4,\\
1{-}a_1{-}a_2{+}a_1a_3{-}a_1a_3a_4{+}a_1^2a_3a_4\}.
\end{split}
\end{equation*}
which we have checked to be spurious at least through weight $8$.  It remains an important open question if these new letters are spurious to arbitrary $L$, but we emphasize that it is already interesting that our estimation does not grow with $L$.

\subsection{More ladder integrals, $D_4, D_5, D_6$ cluster algebras and universality}

\begin{figure}[htbp]
    \centering
    \begin{tikzpicture}[baseline={([yshift=-.5ex]current bounding box.center)},scale=0.18]
        \draw[black,thick] (0,0)--(0,4)--(3.46,6)--(6.93,4)--(6.93,0)--(3.46,-2)--cycle;
        \draw[black,thick] (6.93,4)--(8.66,5);
        \draw[black,thick] (0,4)--(-1.73,5);
        \draw[black,thick] (3.46,-2)--(3.46,-4);
        \filldraw[black] (-1.73,5) node[anchor=south east] {{$1$}};
        \filldraw[black] (2.26,7.73) node[anchor=south] {{$2$}};
        \filldraw[black] (4.66,7.73) node[anchor=south] {{$3$}};
        \filldraw[black] (8.66,5) node[anchor=south west] {{$4$}};
        \filldraw[black] (8.93,0) node[anchor=west] {{$5$}};
        \filldraw[black] (7.93,-1.73) node[anchor=north] {{$6$}};
        \filldraw[black] (3.46,-4) node[anchor=north] {{$7$}};
        \filldraw[black] (-1,-1.73) node[anchor=north] {{$8$}};
        \filldraw[black] (-2,0) node[anchor=east] {{$9$}};
        \draw[black,thick] (2.46,7.73)--(3.46,6)--(4.46,7.73);
        \draw[black,thick] (-2,0)--(0,0)--(-1,-1.73);
        \draw[black,thick] (8.93,0)--(6.93,0)--(7.93,-1.73);
        \filldraw[black] (3.46,2) node[anchor=center] {{6D}};
    \end{tikzpicture}
    \begin{tikzpicture}[baseline={([yshift=-.5ex]current bounding box.center)},scale=0.18]
        \draw[black,thick] (0,0)--(0,4)--(3.46,6)--(6.93,4)--(6.93,0)--(3.46,-2)--cycle;
        \draw[black,thick] (6.93,4)--(8.66,5);
        \draw[black,thick] (0,4)--(-1.73,5);
        \draw[black,thick] (3.46,-2)--(3.46,-4);
        \filldraw[black] (-1.73,5) node[anchor=south east] {{$1$}};
        \filldraw[black] (2.26,7.73) node[anchor=south] {{$2$}};
        \filldraw[black] (4.66,7.73) node[anchor=south] {{$3$}};
        \filldraw[black] (8.66,5) node[anchor=south west] {{$4$}};
        \filldraw[black] (8.93,0) node[anchor=west] {{$5$}};
        \filldraw[black] (7.93,-1.73) node[anchor=north] {{$6$}};
        \filldraw[black] (3.46,-4) node[anchor=north] {{$7$}};
        \filldraw[black] (-1.2,-1.53) node[anchor=north east] {{$8$}};
        \draw[black,thick] (2.46,7.73)--(3.46,6)--(4.46,7.73);
        \draw[black,thick] (0,0)--(-1.2,-1.53);
        \draw[black,thick] (8.93,0)--(6.93,0)--(7.93,-1.73);
        \filldraw[black] (3.46,2) node[anchor=center] {{6D}};
    \end{tikzpicture}
    \begin{tikzpicture}[baseline={([yshift=-.5ex]current bounding box.center)},scale=0.18]
        \draw[black,thick] (0,0)--(0,4)--(3.46,6)--(6.93,4)--(6.93,0)--(3.46,-2)--cycle;
        \draw[black,thick] (6.93,4)--(8.66,5);
        \draw[black,thick] (0,4)--(-1.73,5);
        \draw[black,thick] (3.46,-2)--(3.46,-4);
        \filldraw[black] (-1.73,5) node[anchor=south east] {{$1$}};
        \filldraw[black] (2.26,7.73) node[anchor=south] {{$2$}};
        \filldraw[black] (4.66,7.73) node[anchor=south] {{$3$}};
        \filldraw[black] (8.66,5) node[anchor=south west] {{$4$}};
        \filldraw[black] (8.13,-1.53) node[anchor=north west] {{$5$}};
        \filldraw[black] (3.46,-4) node[anchor=north] {{$6$}};
        \filldraw[black] (-1.2,-1.53) node[anchor=north east] {{$7$}};
        \draw[black,thick] (2.46,7.73)--(3.46,6)--(4.46,7.73);
        \draw[black,thick] (0,0)--(-1.2,-1.53);
        \draw[black,thick] (6.93,0)--(8.13,-1.53);
        \filldraw[black] (3.46,2) node[anchor=center] {{6D}};
    \end{tikzpicture}
    \caption{Three-mass-easy $6D$ hexagon and its degenerations}\label{fig3}
\end{figure}
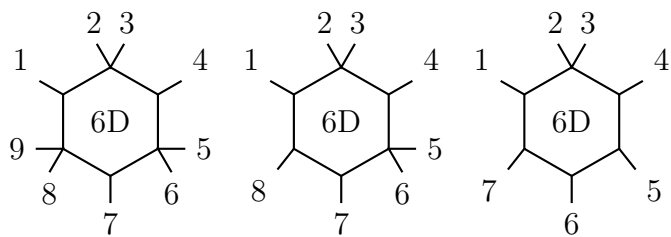

So far we have studied $I_{\rm pb}^{(L)}$ and $I_{\rm dp}^{(L)}$ using our recursion, and it is natural to try and apply it to more examples and to see if the alphabets are related to cluster algebras. We expect it to be a general phenomenon for a large class of Feynman integrals, especially for those referred to as generalized penta-ladder integrals~\cite{He:2020uxy}. Once we know the one-loop case, we can apply the recursion to obtain symbol at higher-loops, and we conjecture that the alphabet will stay invariant starting a certain order. In the first version we have studied a mathematical experiment by applying the recursion relation to (weight-$3$) hexagons in 6d, including one-mass, two-mass-easy and three-mass cases. We have found that by applying recursion relation to these three hexagon integrals: $I^{\rm 6d}_{\rm 3me}$~\cite{DelDuca:2011wh}, $I^{\rm 6d}_{\rm 2me}$, and $I_{\rm 1m}^{6d}$~\cite{DelDuca:2011jm, Spradlin:2011wp} (see figure.~\ref{fig3}), the three new series of functions have alphabet of $D_4$, $D_5$ and a subset of $D_6$ respectively. 


We have not found any physical interpretation of these series of functions; given the results of~\cite{Bourjaily:2018aeq}, it is natural to wonder if those double-penta-ladder integrals (for $n=7,8,9$), which share the same kinematics as these 6d hexagons, also have alphabet of $D_4, D_5, D_6$ respectively. Let us first consider the most general kinematics, the three-mass-easy (nine-point) hexagon (which does not involve any square root in 4d), and it depends on the following $6$ cross-ratios:
\begin{align*}
v_1=\frac{x_{17}^2 x_{25}^2}{x_{15}^2 x_{27}^2}
,\ 
v_2=\frac{x_{14}^2 x_{58}^2}{x_{15}^2 x_{48}^2}
,\ 
v_3=\frac{x_{28}^2 x_{47}^2}{x_{27}^2 x_{48}^2}
,\  
v_4=\frac{x_{15}^2 x_{24}^2}{x_{14}^2 x_{25}^2}
,\ 
v_5=\frac{x_{48}^2 x_{57}^2}{x_{47}^2 x_{58}^2}
,\ 
v_6=\frac{x_{18}^2 x_{27}^2}{x_{17}^2 x_{28}^2}.
\end{align*}
We can parametrize the $9$ momentum twistors as
\begin{align}
{\bf Z}=    \left(
\begin{array}{ccccccccc}
 0 & 1 & 1 & 0 & 1 & 1 & 0 & 1 & 1 \\
 1 & 0 & 1 & 0 & a_5 & 0 & 0 & a_4 & 0 \\
 0 & a_1 & 0 & 1 & 0 & 1 & 0 & a_6 & 0 \\
 0 & a_2 & 0 & 0 & a_3 & 0 & 1 & 0 & 1 \\
\end{array}
\right)\,.
\end{align}
In terms of these variables, the cross-ratios $v_i$ for $i=1,\cdots, 6$ read:
\begin{align*}
&v_1=\frac{a_2-a_3}{\left(a_1-1\right) \left(a_3-1\right)},\ v_2= -\frac{a_4-a_5}{\left(a_3-1\right) \left(a_4-1\right)},\ v_3= -\frac{a_1-a_6}{\left(a_1-1\right) \left(a_4-1\right)},\\
& v_4= -\frac{a_2 \left(a_3-1\right)}{a_2-a_3},\ v_5=\frac{\left(a_4-1\right) a_5}{a_4-a_5},\ v_6=\frac{\left(a_1-1\right) a_6}{a_1-a_6}  
\end{align*}

As shown in~\cite{Bourjaily:2018aeq}, there are four natural choices of the numerators, while in our formalism, we fix the numerator of the right-most pentagon to be the wavy line $N_1=\bar{1} \cap \bar{4}$. Then for the left-most pentagon, we have two choices
\begin{equation}
\begin{split}
&N_2^1=\biggl[((45(791)\cap(14))\cap\bar7)-((91(745)\cap(14))\cap\bar7)\biggr]\\
&N_2^2=\biggl[(7(14(91)\cap\bar7)\cap(45))-(7(14(45)\cap\bar7)\cap(91))\biggr]
\end{split}
\end{equation}
Since our recursion is derived using the right-most pentagon, it turns out to be independent of the choice of numerators $N_2$; we collectively denote such integrals as $F_{\rm 3me}^{(L)}$ (the source at $L=1$ differs by the choice of numerators), and the recursion universally reads: 
\begin{equation}\label{3merecursion}
    \begin{split}
        F^{(L+\frac12)}_{\rm 3me}(v_1,\cdots,v_6)&=\int {\rm d}\log\frac{t{+}1}{t}F^{(L)}_{\rm 3me}\biggl(\frac{v_1(t{+}v_4)}{t{+}v_1 v_4},v_2,\frac{v_3(t{+}v_1 v_4 v_6)}{t{+}v_1 v_4},\frac{v_4(t{+}1)}{t{+}v_4},v_5,\frac{v_6(t{+}v_1 v_4)}{t{+}v_1 v_4 v_6}\biggr)\\
        F^{(L+1)}_{\rm 3me}(v_1,\cdots,v_6)&=\int {\rm d}\log(s{+}1)F^{(L+\frac12)}_{\rm 3me}\biggl(v_1,\frac{v_2(1{+}s)}{1{+}s v_2},\frac{v_3(1{+}s v_2 v_5)}{1{+}s v_2},\frac{s{+}v_4}{s{+}1},\frac{v_5(1+s v_2)}{1+s v_2 v_5},v_6\biggr)
    \end{split}
\end{equation}
Remarkably for $F^{(L+1)}_{\rm 3me}$ with numerator either $(N_1,N_2^1)$ or $(N_1,N_2^2)$ up to $L=4$, we find $35$ out of $36$ letters which form the alphabet of $D_6$\footnote{At $L=2$ level, $F^{(L+1)}_{\rm 3me}$ with numerator $(N_1,N_2^1)$ have $33$ letters out of the $35$. While at $L=3$ level, all $35$ letters appear.}. For example we can use  the following change of variables to arrive at $d=6$ case of \eqref{AD} but with the letter $1+z_4$ missing:
\begin{equation}\label{D6re}
a_1{=}\frac{z_1(1{+}z_5)}{z_1{-}z_5},\  a_2{=}\frac{z_5(z_1{-}z_2)}{z_2(z_1{-}z_5)},\  a_3{=}\frac{z_5(z_3{-}z_2)}{z_2(z_3{-}z_5)},\  a_4{=}\frac{z_2{-}z_5}{z_4{-}z_5},\ a_5{=}\frac{z_2{-}z_5}{z_3{-}z_5},\  a_6{=}\frac{z_4{+}z_5z_6}{z_4{-}z_5}
\end{equation}
It would be interesting to see if the missing letter appears at higher loops.

Now we move to various degenerations. As pointed out in \cite{Bourjaily:2018aeq}, $F^{(L+1)}_{\rm 3me}$ does not have smooth limit $u_5\rightarrow0$ or $u_6\rightarrow0$. The only degeneration we can consider is $u_4\rightarrow0$, {\it i.e.} collinear limit $3\rightarrow2$ or limit $a_2\rightarrow0$ in $a$s variables, after which we get {\bf octagon A}. Up to $L=4$, its alphabet has $24$ letters, which forms $D_5$ with one letter missing as well. To see $D_5$, we can follow the same procedure and find the ${\bf u}$ variables, or alternatively a degeneration $z_1\to z_2$ from \eqref{D6re}. The missing one of $D_5$ alphabet is then the degeneration image of $1+z_4$.

Moreover, we can consider {\bf octagon C} in Fig. \ref{fig2}, which shares the same kinematics as the two-mass-easy hexagon but has different massive corners as {\bf octagon A}. There are also two natural choices of the numerator $N_2$, after fixing $N_1=\bar1\cap\bar4$:
\begin{equation}
\begin{split}
&N_2^1=\biggl[((67)\cap((781)\cap(14)45)8)-((67)\cap((745)\cap(14)81)8)\biggr]\\
&N_2^2=\biggl[(7(81)\cap((14)\cap(458)67))-(7(81)\cap((14)\cap(678)45))\biggr]
\end{split}
\end{equation}
Explicit computation shows that up to $L=3$, its alphabet has $25$ letters which forms the full $D_5$, corresponding to the degeneration $z_4\to-z_5z_6$ of $D_6$ alphabet. Finally, when taking limit $z_1\to z_2$ furthermore, we obtain {\bf heptagon B} with $16$ letters, forming a full $D_4$ alphabet.

\section{Conclusion and Discussions}

In this paper, motivated by~\cite{Chicherin:2020umh} we have studied relations between Feynman integrals and cluster algebras using the recursive ${\rm d}\log$ forms~\cite{He:2020uxy}. Our main examples are the penta-box-ladder integral $I_{\rm pb}^{(L)}$ which has an alphabet of $D_3 \simeq A_3$ and seven-point double-penta-ladder integral $I_{\rm dp}^{(L)}$ which has an alphabet of $D_4$. We have also found that such $D_n$-type cluster algebras seem to be rather universal for (double-penta) ladder integrals: various such integrals with one-mass, two-mass-easy and three-mass-easy hexagon kinematics are associated with cluster algebras $D_4$, $D_5$ and $D_6$ respectively, which is independent of details of the integrals. We have identified the ${\bf u}$ variables of cluster configuration space for a gauge-invariant description. 

The most pressing open question is to understand why alphabets of such ladder integrals can be identified with cluster algebras. For $n=6,7$ cases, it is not surprising that the alphabet of such integrals can be embedded into $A_3$ and $E_6$~\footnote{For example, by taking collinear limits of generic double-pentagon integral~\cite{He:2020lcu}, we have computed another heptagon integral (dubbed {\bf heptagon C} in~\cite{Bourjaily:2018aeq}), whose symbol has been bootstrapped in~\cite{Henn:2018cdp}, and as already noticed there it has $38$ letters which is a subset of $E_6$ alphabet for the full amplitude.}, but it would be interesting to understand possible origin for more general cases. Differential equations or our recursions can provide possible explanations: {\it e.g.} if we could exclude spurious letters, our rules for generating alphabet recursively may turn out to be certain automorphism on cluster algebras. Following~\cite{Caron-Huot:2018dsv}, it is possible to resum these ladders~\cite{Li2021, He:2020uxy}, and one may prove the alphabet by carefully analyzing the all-loop result. It is also important to study what are the cluster algebras for more general integrals including divergent ones, which has been achieved for many cases in~\cite{Chicherin:2020umh} using dimensional regularization, and it would be nice to connect our method to differential equations for studying such divergent integrals. 

As we have seen, seven-point $I_{\rm dp}^{(L)}$ can be directly embedded in $G(4,7)/T \sim E_6$, but generally for $n\geq 8$ we have to consider finite sub-algebras of infinite cases. It is particularly difficult to study integrals with algebraic letters: the box-ladder is a trivial example where we have $A_1^2$ with $z$ and $\bar{z}$ variables, but for generic case with square roots, we expect that multiple cluster algebras or even more complicated situations are needed. For example, even for one-loop, {\it e.g.} the generic (twelve-point) chiral octagon with $16$ different square roots~\cite{ArkaniHamed:2010gh}, there are $16$ $A_1^2$ with their own $z, \bar{z}$ variables, but these $32$ variables are constrained, resulting in a $13$-dimensional space. 

A much more non-trivial example is given by generic double-pentagon integrals (whose symbol was computed in~\cite{He:2020lcu}) with $6\times 16$ algebraic letters and $164$ (A-coordinate like) rational letters. To simplify the problem, we consider the double-pentagon with massless corners $1,4,5,8$ (and the generic eight-point double-penta ladder as the natural generalization), which depends on $5$ cross-ratios and only contains a single square root: we have $v_1, v_2, v_4$ given by \eqref{v1v2v3v4} with $7 \to 8$ and $v_3=x_{28}^2 x_{46}^2/x^2_{26} x_{48}^2$, $v_3'=x_{24}^2 x_{68}^2/x^2_{26} x_{48}^2$, and the square root is the one associated with four-mass box of $v_3, v_3'$, $\Delta(2,4,6,8)=\sqrt{(1-v_3-v_3')^2-4 v_3 v_3'}$. For $L=2$ it was dubbed {\bf octagon B} and first computed in~\cite{Bourjaily:2018aeq}; either from the symbol of that result or from our direct computation we see that, in addition to $5$ (multiplicative independent) algebraic letters given in~\cite{He:2020lcu}, there are $22$ rational letters 

\vspace{-3ex}
{\footnotesize\begin{align*} 
\biggl\{&
v_1,\ v_2,\ v_3,\ v_3',\ v_4,\ 1{-}v_1,\ 1{-}v_2,\ 1{-}v_4,\ 1{-}v_1 v_4, \ 1{-}v_2 v_4,\ v_3'{-} v_1 v_4,\ v_3'{-}v_2 v_4,\ v_3'{-}v_1 v_2 v_4, \\
& v_3'{-} v_1 v_4{-}v_3' v_4{+}v_1 v_2 v_4^2, 
v_3'{-}v_1 v_4{-}v_1 v_3' v_4{+}v_1^2 v_4^2{+}v_1 v_4 v_3, 
(v_1 \leftrightarrow v_2) \\[1.5ex]
& \frac{1{-}x_+}{1{-}x_-},\ \frac{v_4{-}x_+}{v_4{-}x_-},\ \frac{1{-}v_1{-}v_2{+}v_1 v_2 x_+}{1{-}v_1{-}v_2{+} v_1 v_2 x_-},\ \frac{v_1^{-1}{-}y_+}{v_1^{-1}{-}y_-},\frac{v_2^{-1}{-}y_+}{v_2^{-1}{-}y_-}\biggr\}
\end{align*}}%
where we have abbreviated two letters with $(v_1 \leftrightarrow v_2)$ on the second line (note the integral is symmetric in $v_1 \leftrightarrow v_2)$, and on the third line we have the combinations
\begin{align*}x_\pm&=\frac{-v_3'-v_4+v_1 v_4+v_2v_4-v_3'v_4+v_1 v_2 v_4^2+v_3v_4\pm\sqrt{\Delta_8}}{2(v_1 v_2 v_4-v_3')},\\
y_\pm&=\frac{-v_4(-1+v_1+v_2-v_3'+v_3)+(-v_3'+v_1 v_2 v_4)x_\pm}{v_3'(1-v_1-v_2+v_1 v_2 v_4)-v_1 v_2 v_3 v_4},
\end{align*}
with $\Delta_8=(v_3'+v_4-v_1 v_4-v_2v_4-v_3'v_4+v_1 v_2 v_4^2+v_3v_4)^2-4(1-v_1)(1-v_2)v_4^2v_3$, which become rational in terms of momentum twistors. In addition, by a change of variables from $2$ of the parameters (simply related to $v_3$, $v_3'$) to $z$, $\bar{z}$ of $\Delta_{1,4,5,8}$ (see \eqref{zzb}), we find the the square roots for algebraic letters disappear as well. It remains an interesting open question to find cluster algebra interpretation of this alphabet, and to generalize it to higher loops. One possibility is to try embedding the alphabet in the octagon alphabet from tropical $G_+(4,8)$~\cite{Drummond:2019cxm, Henke:2019hve, Arkani-Hamed:2019rds}. This example may tell us what to expect in general for alphabet with algebraic letters. 

We have only briefly discussed cluster adjacency/extended Steinmann relations for such ladder integrals, and it would be interesting to study the (reduced) cluster function spaces for $D_3$ and $D_4$, as well as how to locate ladder integrals there. It is plausible that by imposing constraints from co-products of the integrals, which follow from our ${\rm d} \log$ form or equivalently differential equations (plus the symmetry under exchange of $v_1, v_2$ and smooth collinear limits), they can be bootstrapped to relatively high orders. Perhaps the most intriguing question is (see \cite{Caron-Huot:2018dsv}): where can we see all these cluster-algebra structures in the resummed, non-pertubative result?

\section*{Acknowledgement}
We are grateful to Nima Arkani-Hamed, Johannes Henn and Georgios Papathanasiou for inspiring discussions/comments. We also thank Yichao Tang and Chi Zhang for collaborations on related projects. This research is supported in part by National Natural Science Foundation of China under Grant No. 11935013 and Peng Huanwu center under Grant No. 12047503.

\appendix

\section{Explicit result for three-loop penta-box ladder}
Here we present $I^{(L)}_{\rm pb}$ for $L=3$ as a linear 
combination of Goncharov polylogarithms with weight $6$. Recall that such a weight-$w$ $G$-function is defined by $w$-fold iterated integrals~\cite{goncharov2005galois}
\begin{equation}
    G(a_{1},\ldots,a_{w};z):=\int_{0}^{z}\frac{\dif t}{t-a_{1}}\, G(a_{2},\ldots,a_{w};t), \label{Gpolylot} 
\end{equation} 
with the starting point $G(;z):=1$. We have not attempted to simplify the result directly obtained from our code (except that the code automatically recognize some classical polylogs in the expression). The weight-$6$ function reads:

\vspace{-2.5ex}
{\footnotesize
\begin{align*}
& G(0,0,1,0,x,1;z)+G(0,0,1,x,0,1;z)-2 G(0,0,1,x,x,1;z)+G(0,0,x,0,0,1;z)-\\
& 2 G(0,0,x,0,x,1;z)-G(0,0,x,1,x,1;z)- 2 G(0,0,x,x,0,1;z)+4 G(0,0,x,x,x,1;z)-\\
& G(0,0,y,0,z,y;y z)+G(0,0,y,x y,y,z;y z)+G(0,0,y,x y,z,y;y z)-G(0,0,y,z,0,y;y z)-\\
& G(0,0,y,z,y,z;y z)+G(0,0,y,z,x y,y;y z)-2 G(0,0,x y,x y,z,y;y z)+G(0,0,x y,0,z,y;y z)+\\
& G(0,0,x y,y,0,z;y z)+G(0,0,x y,y,z,y;y z)-2 G(0,0,x y,x y,y,z;y z)+G(0,0,x y,0,y,z;y z)+\\
& G(0,0,x y,z,0,y;y z)+G(0,0,x y,z,y,z;y z)-G(0,0,x y,z,x y,y;y z)-G(0,0,z,0,0,y;y z)-\\
& G(0,0,z,0,y,z;y z)+G(0,0,z,0,x y,y;y z)-G(0,0,z,y,0,z;y z)+G(0,0,z,y,x y,y;y z)-\\
& G(0,0,z,y,z,y;y z)+G(0,0,z,x y,0,y;y z)+G(0,0,z,x y,y,z;y z)-2 G(0,0,z,x y,x y,y;y z)+\\
& G(0,0,z,x y,z,y;y z)+G(0,1,0,1,x,1;z)+G(0,1,0,x,0,1;z)-2 G(0,1,0,x,x,1;z)-\\
& G(0,y,0,y,z,y;y z)+G(0,y,0,x y,y,z;y z)+G(0,y,0,x y,z,y;y z)-G(0,y,0,z,0,y;y z)-\\
& G(0,y,0,z,y,z;y z)+G(0,y,0,z,x y,y;y z)+ \log (1-x)\bigl(-G(0,0,1,0,x;z)-G(0,0,1,x,1;z)+\\
& 2 G(0,0,1,x,x;z)-G(0,0,x,0,1;z)+2 G(0,0,x,0,x;z)+G(0,0,x,1,x;z)+2 G(0,0,x,x,1;z)-\\
& 4 G(0,0,x,x,x;z)+G(0,0,y,0,z;y z)-G(0,0,y,x y,z;y z)+G(0,0,y,z,y;y z)-\\
& G(0,0,y,z,x y;y z)-G(0,0,x y,0,z;y z)-G(0,0,x y,y,z;y z)+2 G(0,0,x y,x y,z;y z)-\\
& G(0,0,x y,z,y;y z)+G(0,0,x y,z,x y;y z)+G(0,0,z,0,y;y z)-G(0,0,z,0,x y;y z)-\\
& G(0,0,z,y,x y;y z)+G(0,0,z,y,z;y z)-G(0,0,z,x y,y;y z)+2 G(0,0,z,x y,x y;y z)-\\
& G(0,0,z,x y,z;y z)-G(0,1,0,1,x;z)-G(0,1,0,x,1;z)+2 G(0,1,0,x,x;z)+\\
& G(0,y,0,y,z;y z)-G(0,y,0,x y,z;y z)+G(0,y,0,z,y;y z)-G(0,y,0,z,x y;y z)\bigr)+\\
& \operatorname{Li}_2(x)\bigl(-G(0,0,1,x;z)-G(0,0,x,1;z)+2 G(0,0,x,x;z)+G(0,0,y,z;y z)-G(0,0,x y,z;y z)+\\
& G(0,0,z,y;y z)-G(0,0,z,x y;y z)-G(0,1,0,x;z)+G(0,y,0,z;y z)\bigr) +\operatorname{Li}_3(x)\left(\operatorname{Li}_3(y)-\operatorname{Li}_3\left(\frac{z}{x}\right)\right) 
\end{align*}}
with $x=-z_2,y=-z_3,z=-z_1$.

\bibliographystyle{utphys}
\bibliography{main}

\providecommand{\noopsort}[1]{}\providecommand{\singleletter}[1]{#1}%
\providecommand{\href}[2]{#2}\begingroup\raggedright\begin{thebibliography}{10}

\bibitem{ArkaniHamed:2010kv}
N.~Arkani-Hamed, J.~L. Bourjaily, F.~Cachazo, S.~Caron-Huot, and J.~Trnka,
  ``{The All-Loop Integrand For Scattering Amplitudes in Planar N=4 SYM},''
  \href{http://dx.doi.org/10.1007/JHEP01(2011)041}{{\em JHEP} {\bfseries 01}
  (2011) 041},
\href{http://arxiv.org/abs/1008.2958}{{\ttfamily arXiv:1008.2958 [hep-th]}}.

\bibitem{Arkani-Hamed:2016byb}
N.~Arkani-Hamed, J.~L. Bourjaily, F.~Cachazo, A.~B. Goncharov, A.~Postnikov,
  and J.~Trnka, \href{http://dx.doi.org/10.1017/CBO9781316091548}{{\em
  {Grassmannian Geometry of Scattering Amplitudes}}}.
\newblock Cambridge University Press, 4, 2016.
\newblock \href{http://arxiv.org/abs/1212.5605}{{\ttfamily arXiv:1212.5605
  [hep-th]}}.

\bibitem{Arkani-Hamed:2013jha}
N.~Arkani-Hamed and J.~Trnka, ``{The Amplituhedron},''
  \href{http://dx.doi.org/10.1007/JHEP10(2014)030}{{\em JHEP} {\bfseries 10}
  (2014) 030},
\href{http://arxiv.org/abs/1312.2007}{{\ttfamily arXiv:1312.2007 [hep-th]}}.

\bibitem{Dixon:2011pw}
L.~J. Dixon, J.~M. Drummond, and J.~M. Henn, ``{Bootstrapping the three-loop
  hexagon},'' \href{http://dx.doi.org/10.1007/JHEP11(2011)023}{{\em JHEP}
  (2011) 023},
\href{http://arxiv.org/abs/1108.4461}{{\ttfamily arXiv:1108.4461 [hep-th]}}.

\bibitem{Dixon:2014xca}
L.~J. Dixon, J.~M. Drummond, C.~Duhr, M.~von Hippel, and J.~Pennington,
  ``{Bootstrapping six-gluon scattering in planar N=4 super-Yang-Mills
  theory},'' \href{http://dx.doi.org/10.22323/1.211.0077}{{\em PoS} {\bfseries
  LL2014} (2014) 077},
\href{http://arxiv.org/abs/1407.4724}{{\ttfamily arXiv:1407.4724 [hep-th]}}.

\bibitem{Dixon:2014iba}
L.~J. Dixon and M.~von Hippel, ``{Bootstrapping an NMHV amplitude through three
  loops},'' \href{http://dx.doi.org/10.1007/JHEP10(2014)065}{{\em JHEP}
  {\bfseries 10} (2014) 065},
\href{http://arxiv.org/abs/1408.1505}{{\ttfamily arXiv:1408.1505 [hep-th]}}.

\bibitem{Drummond:2014ffa}
J.~M. Drummond, G.~Papathanasiou, and M.~Spradlin, ``{A Symbol of Uniqueness:
  The Cluster Bootstrap for the 3-Loop MHV Heptagon},''
  \href{http://dx.doi.org/10.1007/JHEP03(2015)072}{{\em JHEP} {\bfseries 03}
  (2015) 072},
\href{http://arxiv.org/abs/1412.3763}{{\ttfamily arXiv:1412.3763 [hep-th]}}.

\bibitem{Dixon:2015iva}
L.~J. Dixon, M.~von Hippel, and A.~J. McLeod, ``{The four-loop six-gluon NMHV
  ratio function},'' \href{http://dx.doi.org/10.1007/JHEP01(2016)053}{{\em
  JHEP} {\bfseries 01} (2016) 053},
\href{http://arxiv.org/abs/1509.08127}{{\ttfamily arXiv:1509.08127 [hep-th]}}.

\bibitem{Caron-Huot:2016owq}
S.~Caron-Huot, L.~J. Dixon, A.~McLeod, and M.~von Hippel, ``{Bootstrapping a
  Five-Loop Amplitude Using Steinmann Relations},''
  \href{http://dx.doi.org/10.1103/PhysRevLett.117.241601}{{\em Phys. Rev.
  Lett.} {\bfseries 117} no.~24, (2016) 241601},
\href{http://arxiv.org/abs/1609.00669}{{\ttfamily arXiv:1609.00669 [hep-th]}}.

\bibitem{Dixon:2016nkn}
L.~J. Dixon, J.~Drummond, T.~Harrington, A.~J. McLeod, G.~Papathanasiou, and
  M.~Spradlin, ``{Heptagons from the Steinmann Cluster Bootstrap},''
  \href{http://dx.doi.org/10.1007/JHEP02(2017)137}{{\em JHEP} {\bfseries 02}
  (2017) 137},
\href{http://arxiv.org/abs/1612.08976}{{\ttfamily arXiv:1612.08976 [hep-th]}}.

\bibitem{Drummond:2018caf}
J.~Drummond, J.~Foster, {\"{O}}.~G{\"{u}}rdo{\u{g}}an, and G.~Papathanasiou,
  ``{Cluster adjacency and the four-loop NMHV heptagon},''
  \href{http://dx.doi.org/10.1007/JHEP03(2019)087}{{\em JHEP} {\bfseries 03}
  (2019) 087},
\href{http://arxiv.org/abs/1812.04640}{{\ttfamily arXiv:1812.04640 [hep-th]}}.

\bibitem{Caron-Huot:2019vjl}
S.~Caron-Huot, L.~J. Dixon, F.~Dulat, M.~von Hippel, A.~J. McLeod, and
  G.~Papathanasiou, ``{Six-Gluon amplitudes in planar $ \mathcal{N} $ = 4
  super-Yang-Mills theory at six and seven loops},''
  \href{http://dx.doi.org/10.1007/JHEP08(2019)016}{{\em JHEP} {\bfseries 08}
  (2019) 016},
\href{http://arxiv.org/abs/1903.10890}{{\ttfamily arXiv:1903.10890 [hep-th]}}.

\bibitem{Caron-Huot:2019bsq}
S.~Caron-Huot, L.~J. Dixon, F.~Dulat, M.~von Hippel, A.~J. McLeod, and
  G.~Papathanasiou, ``{The Cosmic Galois Group and Extended Steinmann Relations
  for Planar $\mathcal{N} = 4$ SYM Amplitudes},''
  \href{http://dx.doi.org/10.1007/JHEP09(2019)061}{{\em JHEP} {\bfseries 09}
  (2019) 061},
\href{http://arxiv.org/abs/1906.07116}{{\ttfamily arXiv:1906.07116 [hep-th]}}.

\bibitem{Dixon:2020cnr}
L.~J. Dixon and Y.-T. Liu, ``{Lifting Heptagon Symbols to Functions},''
  \href{http://dx.doi.org/10.1007/JHEP10(2020)031}{{\em JHEP} {\bfseries 10}
  (2020) 031}, \href{http://arxiv.org/abs/2007.12966}{{\ttfamily
  arXiv:2007.12966 [hep-th]}}.

\bibitem{Caron-Huot:2020bkp}
S.~Caron-Huot, L.~J. Dixon, J.~M. Drummond, F.~Dulat, J.~Foster,
  O.~G\"urdo\u{g}an, M.~von Hippel, A.~J. McLeod, and G.~Papathanasiou, ``{The
  Steinmann Cluster Bootstrap for $N$ = 4 Super Yang-Mills Amplitudes},''
  \href{http://dx.doi.org/10.22323/1.376.0003}{{\em PoS} {\bfseries CORFU2019}
  (2020) 003}, \href{http://arxiv.org/abs/2005.06735}{{\ttfamily
  arXiv:2005.06735 [hep-th]}}.

\bibitem{Golden:2013xva}
J.~Golden, A.~B. Goncharov, M.~Spradlin, C.~Vergu, and A.~Volovich, ``{Motivic
  Amplitudes and Cluster Coordinates},''
  \href{http://dx.doi.org/10.1007/JHEP01(2014)091}{{\em JHEP} {\bfseries 01}
  (2014) 091},
\href{http://arxiv.org/abs/1305.1617}{{\ttfamily arXiv:1305.1617 [hep-th]}}.

\bibitem{Zhang:2019vnm}
S.~He, Z.~Li, and C.~Zhang, ``{Two-loop Octagons, Algebraic Letters and
  $\bar{Q}$ Equations},''
  \href{http://dx.doi.org/10.1103/PhysRevD.101.061701}{{\em Phys. Rev. D}
  {\bfseries 101} no.~6, (2020) 061701},
  \href{http://arxiv.org/abs/1911.01290}{{\ttfamily arXiv:1911.01290
  [hep-th]}}.

\bibitem{He:2020vob}
S.~He, Z.~Li, and C.~Zhang, ``{The symbol and alphabet of two-loop NMHV
  amplitudes from $\bar{Q}$ equations},''
  \href{http://arxiv.org/abs/2009.11471}{{\ttfamily arXiv:2009.11471
  [hep-th]}}.

\bibitem{CaronHuot:2011kk}
S.~Caron-Huot and S.~He, ``{Jumpstarting the All-Loop S-Matrix of Planar N=4
  Super Yang-Mills},'' \href{http://dx.doi.org/10.1007/JHEP07(2012)174}{{\em
  JHEP} {\bfseries 07} (2012) 174},
\href{http://arxiv.org/abs/1112.1060}{{\ttfamily arXiv:1112.1060 [hep-th]}}.

\bibitem{Drummond:2019cxm}
J.~Drummond, J.~Foster, O.~G\"urdogan, and C.~Kalousios, ``{Algebraic
  singularities of scattering amplitudes from tropical geometry},''
  \href{http://arxiv.org/abs/1912.08217}{{\ttfamily arXiv:1912.08217
  [hep-th]}}.

\bibitem{Henke:2019hve}
N.~Henke and G.~Papathanasiou, ``{How tropical are seven- and eight-particle
  amplitudes?},'' \href{http://dx.doi.org/10.1007/JHEP08(2020)005}{{\em JHEP}
  {\bfseries 08} (2020) 005}, \href{http://arxiv.org/abs/1912.08254}{{\ttfamily
  arXiv:1912.08254 [hep-th]}}.

\bibitem{Arkani-Hamed:2019rds}
N.~Arkani-Hamed, T.~Lam, and M.~Spradlin, ``{Non-perturbative geometries for
  planar $\mathcal{N}=4$ SYM amplitudes},''
  \href{http://arxiv.org/abs/1912.08222}{{\ttfamily arXiv:1912.08222
  [hep-th]}}.

\bibitem{Herderschee:2021dez}
A.~Herderschee, ``{Algebraic branch points at all loop orders from positive
  kinematics and wall crossing},''
  \href{http://arxiv.org/abs/2102.03611}{{\ttfamily arXiv:2102.03611
  [hep-th]}}.

\bibitem{Mago:2020kmp}
J.~Mago, A.~Schreiber, M.~Spradlin, and A.~Volovich, ``{Symbol alphabets from
  plabic graphs},'' \href{http://dx.doi.org/10.1007/JHEP10(2020)128}{{\em JHEP}
  {\bfseries 10} (2020) 128}, \href{http://arxiv.org/abs/2007.00646}{{\ttfamily
  arXiv:2007.00646 [hep-th]}}.

\bibitem{He:2020uhb}
S.~He and Z.~Li, ``{A Note on Letters of Yangian Invariants},''
  \href{http://dx.doi.org/10.1007/JHEP02(2021)155}{{\em JHEP} {\bfseries 02}
  (2021) 155}, \href{http://arxiv.org/abs/2007.01574}{{\ttfamily
  arXiv:2007.01574 [hep-th]}}.

\bibitem{Mago:2020nuv}
J.~Mago, A.~Schreiber, M.~Spradlin, A.~Yelleshpur~Srikant, and A.~Volovich,
  ``{Symbol Alphabets from Plabic Graphs II: Rational Letters},''
  \href{http://arxiv.org/abs/2012.15812}{{\ttfamily arXiv:2012.15812
  [hep-th]}}.

\bibitem{Goncharov:2010jf}
A.~B. Goncharov, M.~Spradlin, C.~Vergu, and A.~Volovich, ``{Classical
  Polylogarithms for Amplitudes and Wilson Loops},''
  \href{http://dx.doi.org/10.1103/PhysRevLett.105.151605}{{\em Phys. Rev.
  Lett.} {\bfseries 105} (2010) 151605},
\href{http://arxiv.org/abs/1006.5703}{{\ttfamily arXiv:1006.5703 [hep-th]}}.

\bibitem{Duhr:2011zq}
C.~Duhr, H.~Gangl, and J.~R. Rhodes, ``{From polygons and symbols to
  polylogarithmic functions},''
  \href{http://dx.doi.org/10.1007/JHEP10(2012)075}{{\em JHEP} {\bfseries 10}
  (2012) 075},
\href{http://arxiv.org/abs/1110.0458}{{\ttfamily arXiv:1110.0458 [math-ph]}}.

\bibitem{ArkaniHamed:2010gh}
N.~Arkani-Hamed, J.~L. Bourjaily, F.~Cachazo, and J.~Trnka, ``{Local Integrals
  for Planar Scattering Amplitudes},''
  \href{http://dx.doi.org/10.1007/JHEP06(2012)125}{{\em JHEP} {\bfseries 06}
  (2012) 125},
\href{http://arxiv.org/abs/1012.6032}{{\ttfamily arXiv:1012.6032 [hep-th]}}.

\bibitem{Drummond:2010cz}
J.~M. Drummond, J.~M. Henn, and J.~Trnka, ``{New differential equations for
  on-shell loop integrals},''
  \href{http://dx.doi.org/10.1007/JHEP04(2011)083}{{\em JHEP} {\bfseries 04}
  (2011) 083}, \href{http://arxiv.org/abs/1010.3679}{{\ttfamily arXiv:1010.3679
  [hep-th]}}.

\bibitem{Henn:2013pwa}
J.~M. Henn, ``{Multiloop integrals in dimensional regularization made
  simple},'' \href{http://dx.doi.org/10.1103/PhysRevLett.110.251601}{{\em Phys.
  Rev. Lett.} {\bfseries 110} (2013) 251601},
  \href{http://arxiv.org/abs/1304.1806}{{\ttfamily arXiv:1304.1806 [hep-th]}}.

\bibitem{Henn:2014qga}
J.~M. Henn, ``{Lectures on differential equations for Feynman integrals},''
  \href{http://dx.doi.org/10.1088/1751-8113/48/15/153001}{{\em J. Phys.}
  {\bfseries A48} (2015) 153001},
\href{http://arxiv.org/abs/1412.2296}{{\ttfamily arXiv:1412.2296 [hep-ph]}}.

\bibitem{CaronHuot:2011ky}
S.~Caron-Huot, ``{Superconformal symmetry and two-loop amplitudes in planar N=4
  super Yang-Mills},'' \href{http://dx.doi.org/10.1007/JHEP12(2011)066}{{\em
  JHEP} {\bfseries 12} (2011) 066},
\href{http://arxiv.org/abs/1105.5606}{{\ttfamily arXiv:1105.5606 [hep-th]}}.

\bibitem{Alday:2007hr}
L.~F. Alday and J.~M. Maldacena, ``{Gluon scattering amplitudes at strong
  coupling},'' \href{http://dx.doi.org/10.1088/1126-6708/2007/06/064}{{\em
  JHEP} {\bfseries 06} (2007) 064},
  \href{http://arxiv.org/abs/0705.0303}{{\ttfamily arXiv:0705.0303 [hep-th]}}.

\bibitem{Alday:2007he}
L.~F. Alday and J.~Maldacena, ``{Comments on gluon scattering amplitudes via
  AdS/CFT},'' \href{http://dx.doi.org/10.1088/1126-6708/2007/11/068}{{\em JHEP}
  {\bfseries 11} (2007) 068}, \href{http://arxiv.org/abs/0710.1060}{{\ttfamily
  arXiv:0710.1060 [hep-th]}}.

\bibitem{Drummond:2007aua}
J.~Drummond, G.~Korchemsky, and E.~Sokatchev, ``{Conformal properties of
  four-gluon planar amplitudes and Wilson loops},''
  \href{http://dx.doi.org/10.1016/j.nuclphysb.2007.11.041}{{\em Nucl. Phys. B}
  {\bfseries 795} (2008) 385--408},
  \href{http://arxiv.org/abs/0707.0243}{{\ttfamily arXiv:0707.0243 [hep-th]}}.

\bibitem{Brandhuber:2007yx}
A.~Brandhuber, P.~Heslop, and G.~Travaglini, ``{MHV amplitudes in N=4 super
  Yang-Mills and Wilson loops},''
  \href{http://dx.doi.org/10.1016/j.nuclphysb.2007.11.002}{{\em Nucl. Phys. B}
  {\bfseries 794} (2008) 231--243},
  \href{http://arxiv.org/abs/0707.1153}{{\ttfamily arXiv:0707.1153 [hep-th]}}.

\bibitem{Mason:2010yk}
L.~Mason and D.~Skinner, ``{The Complete Planar S-matrix of N=4 SYM as a Wilson
  Loop in Twistor Space},''
  \href{http://dx.doi.org/10.1007/JHEP12(2010)018}{{\em JHEP} {\bfseries 12}
  (2010) 018}, \href{http://arxiv.org/abs/1009.2225}{{\ttfamily arXiv:1009.2225
  [hep-th]}}.

\bibitem{CaronHuot:2010ek}
S.~Caron-Huot, ``{Notes on the scattering amplitude / Wilson loop duality},''
  \href{http://dx.doi.org/10.1007/JHEP07(2011)058}{{\em JHEP} {\bfseries 07}
  (2011) 058}, \href{http://arxiv.org/abs/1010.1167}{{\ttfamily arXiv:1010.1167
  [hep-th]}}.

\bibitem{He:2020uxy}
S.~He, Z.~Li, Y.~Tang, and Q.~Yang, ``{The Wilson-loop $d \log$ representation
  for Feynman integrals},'' \href{http://arxiv.org/abs/2012.13094}{{\ttfamily
  arXiv:2012.13094 [hep-th]}}.

\bibitem{He:2020lcu}
S.~He, Z.~Li, Q.~Yang, and C.~Zhang, ``{Feynman Integrals and Scattering
  Amplitudes from Wilson Loops},''
  \href{http://arxiv.org/abs/2012.15042}{{\ttfamily arXiv:2012.15042
  [hep-th]}}.

\bibitem{Caron-Huot:2018dsv}
S.~Caron-Huot, L.~J. Dixon, M.~von Hippel, A.~J. McLeod, and G.~Papathanasiou,
  ``{The Double Pentaladder Integral to All Orders},''
  \href{http://dx.doi.org/10.1007/JHEP07(2018)170}{{\em JHEP} {\bfseries 07}
  (2018) 170},
\href{http://arxiv.org/abs/1806.01361}{{\ttfamily arXiv:1806.01361 [hep-th]}}.

\bibitem{Drummond:2017ssj}
J.~Drummond, J.~Foster, and {\"{O}}.~G{\"{u}}rdo{\u{g}}an, ``{Cluster Adjacency
  Properties of Scattering Amplitudes in $N=4$ Supersymmetric Yang-Mills
  Theory},'' \href{http://dx.doi.org/10.1103/PhysRevLett.120.161601}{{\em Phys.
  Rev. Lett.} {\bfseries 120} no.~16, (2018) 161601},
\href{http://arxiv.org/abs/1710.10953}{{\ttfamily arXiv:1710.10953 [hep-th]}}.

\bibitem{Chicherin:2017dob}
D.~Chicherin, J.~Henn, and V.~Mitev, ``{Bootstrapping pentagon functions},''
  \href{http://dx.doi.org/10.1007/JHEP05(2018)164}{{\em JHEP} {\bfseries 05}
  (2018) 164}, \href{http://arxiv.org/abs/1712.09610}{{\ttfamily
  arXiv:1712.09610 [hep-th]}}.

\bibitem{Henn:2018cdp}
J.~Henn, E.~Herrmann, and J.~Parra-Martinez, ``{Bootstrapping two-loop Feynman
  integrals for planar $ \mathcal{N}=4 $ sYM},''
  \href{http://dx.doi.org/10.1007/JHEP10(2018)059}{{\em JHEP} {\bfseries 10}
  (2018) 059},
\href{http://arxiv.org/abs/1806.06072}{{\ttfamily arXiv:1806.06072 [hep-th]}}.

\bibitem{Dixon:2020bbt}
L.~J. Dixon, A.~J. McLeod, and M.~Wilhelm, ``{A Three-Point Form Factor Through
  Five Loops},'' \href{http://arxiv.org/abs/2012.12286}{{\ttfamily
  arXiv:2012.12286 [hep-th]}}.

\bibitem{Chicherin:2020umh}
D.~Chicherin, J.~M. Henn, and G.~Papathanasiou, ``{Cluster algebras for Feynman
  integrals},'' \href{http://arxiv.org/abs/2012.12285}{{\ttfamily
  arXiv:2012.12285 [hep-th]}}.

\bibitem{Hodges:2009hk}
A.~Hodges, ``{Eliminating spurious poles from gauge-theoretic amplitudes},''
  \href{http://dx.doi.org/10.1007/JHEP05(2013)135}{{\em JHEP} {\bfseries 05}
  (2013) 135},
\href{http://arxiv.org/abs/0905.1473}{{\ttfamily arXiv:0905.1473 [hep-th]}}.

\bibitem{Bourjaily:2018aeq}
J.~L. Bourjaily, A.~J. McLeod, M.~von Hippel, and M.~Wilhelm, ``{Rationalizing
  Loop Integration},'' \href{http://dx.doi.org/10.1007/JHEP08(2018)184}{{\em
  JHEP} {\bfseries 08} (2018) 184},
\href{http://arxiv.org/abs/1805.10281}{{\ttfamily arXiv:1805.10281 [hep-th]}}.

\bibitem{DelDuca:2011wh}
V.~Del~Duca, L.~J. Dixon, J.~M. Drummond, C.~Duhr, J.~M. Henn, and V.~A.
  Smirnov, ``{The one-loop six-dimensional hexagon integral with three massive
  corners},'' \href{http://dx.doi.org/10.1103/PhysRevD.84.045017}{{\em Phys.
  Rev. D} {\bfseries 84} (2011) 045017},
  \href{http://arxiv.org/abs/1105.2011}{{\ttfamily arXiv:1105.2011 [hep-th]}}.

\bibitem{Goncharov:2005sla}
A.~B. Goncharov, ``{Galois symmetries of fundamental groupoids and
  noncommutative geometry},''
  \href{http://dx.doi.org/10.1215/S0012-7094-04-12822-2}{{\em Duke Math. J.}
  {\bfseries 128} (2005) 209},
  \href{http://arxiv.org/abs/math/0208144}{{\ttfamily arXiv:math/0208144}}.

\bibitem{Panzer:2014caa}
E.~Panzer, ``{Algorithms for the symbolic integration of hyperlogarithms with
  applications to Feynman integrals},''
  \href{http://dx.doi.org/10.1016/j.cpc.2014.10.019}{{\em Comput. Phys.
  Commun.} {\bfseries 188} (2015) 148--166},
  \href{http://arxiv.org/abs/1403.3385}{{\ttfamily arXiv:1403.3385 [hep-th]}}.

\bibitem{fomin2002cluster}
S.~Fomin and A.~Zelevinsky, ``Cluster algebras i: foundations,'' {\em Journal
  of the American Mathematical Society} {\bfseries 15} no.~2, (2002) 497--529.

\bibitem{fomin2003cluster}
S.~Fomin and A.~Zelevinsky, ``Cluster algebras ii: Finite type
  classification,'' {\em Inventiones mathematicae} {\bfseries 154} no.~1,
  (2003) 63--121.

\bibitem{berenstein2005cluster}
A.~Berenstein, S.~Fomin, A.~Zelevinsky, {\em et~al.}, ``Cluster algebras iii:
  Upper bounds and double bruhat cells,'' {\em Duke Mathematical Journal}
  {\bfseries 126} no.~1, (2005) 1--52.

\bibitem{fomin2007cluster}
S.~Fomin and A.~Zelevinsky, ``Cluster algebras iv: coefficients,'' {\em
  Compositio Mathematica} {\bfseries 143} no.~1, (2007) 112--164.

\bibitem{Golden:2014xqa}
J.~Golden, M.~F. Paulos, M.~Spradlin, and A.~Volovich, ``{Cluster
  Polylogarithms for Scattering Amplitudes},''
  \href{http://dx.doi.org/10.1088/1751-8113/47/47/474005}{{\em J. Phys.}
  {\bfseries A47} no.~47, (2014) 474005},
\href{http://arxiv.org/abs/1401.6446}{{\ttfamily arXiv:1401.6446 [hep-th]}}.

\bibitem{Parker:2015cia}
D.~Parker, A.~Scherlis, M.~Spradlin, and A.~Volovich, ``{Hedgehog bases for
  A$_{n}$ cluster polylogarithms and an application to six-point amplitudes},''
  \href{http://dx.doi.org/10.1007/JHEP11(2015)136}{{\em JHEP} {\bfseries 11}
  (2015) 136}, \href{http://arxiv.org/abs/1507.01950}{{\ttfamily
  arXiv:1507.01950 [hep-th]}}.

\bibitem{Arkani-Hamed:2019plo}
N.~Arkani-Hamed, S.~He, T.~Lam, and H.~Thomas, ``{Binary Geometries,
  Generalized Particles and Strings, and Cluster Algebras},''
  \href{http://arxiv.org/abs/1912.11764}{{\ttfamily arXiv:1912.11764
  [hep-th]}}.

\bibitem{Arkani-Hamed:2020tuz}
N.~Arkani-Hamed, S.~He, and T.~Lam, ``{Cluster configuration spaces of finite
  type},'' \href{http://arxiv.org/abs/2005.11419}{{\ttfamily arXiv:2005.11419
  [math.AG]}}.

\bibitem{Arkani-Hamed:2019mrd}
N.~Arkani-Hamed, S.~He, and T.~Lam, ``{Stringy canonical forms},''
  \href{http://dx.doi.org/10.1007/JHEP02(2021)069}{{\em JHEP} {\bfseries 02}
  (2021) 069}, \href{http://arxiv.org/abs/1912.08707}{{\ttfamily
  arXiv:1912.08707 [hep-th]}}.

\bibitem{Arkani-Hamed:2017mur}
N.~Arkani-Hamed, Y.~Bai, S.~He, and G.~Yan, ``{Scattering Forms and the
  Positive Geometry of Kinematics, Color and the Worldsheet},''
  \href{http://dx.doi.org/10.1007/JHEP05(2018)096}{{\em JHEP} {\bfseries 05}
  (2018) 096}, \href{http://arxiv.org/abs/1711.09102}{{\ttfamily
  arXiv:1711.09102 [hep-th]}}.

\bibitem{Arkani-Hamed:2019vag}
N.~Arkani-Hamed, S.~He, G.~Salvatori, and H.~Thomas, ``{Causal Diamonds,
  Cluster Polytopes and Scattering Amplitudes},''
  \href{http://arxiv.org/abs/1912.12948}{{\ttfamily arXiv:1912.12948
  [hep-th]}}.

\bibitem{brown2009multiple}
F.~Brown, ``Multiple zeta values and periods of moduli spaces
  $\overline{\mathfrak {m}}_{0, n}$,'' in {\em Annales scientifiques de
  l'{\'E}cole Normale Sup{\'e}rieure}, vol.~42, pp.~371--489.
\newblock 2009.

\bibitem{ussyukina1993exact}
N.~Ussyukina and A.~I. Davydychev, ``Exact results for three-and four-point
  ladder diagrams with an arbitrary number of rungs,'' {\em Physics Letters B}
  {\bfseries 305} no.~1-2, (1993) 136--143.

\bibitem{Caron-Huot:2013vda}
S.~Caron-Huot and S.~He, ``{Three-loop octagons and $n$-gons in maximally
  supersymmetric Yang-Mills theory},''
  \href{http://dx.doi.org/10.1007/JHEP08(2013)101}{{\em JHEP} {\bfseries 08}
  (2013) 101},
\href{http://arxiv.org/abs/1305.2781}{{\ttfamily arXiv:1305.2781 [hep-th]}}.

\bibitem{DelDuca:2011jm}
V.~Del~Duca, C.~Duhr, and V.~A. Smirnov, ``{The One-Loop One-Mass Hexagon
  Integral in D=6 Dimensions},''
  \href{http://dx.doi.org/10.1007/JHEP07(2011)064}{{\em JHEP} {\bfseries 07}
  (2011) 064}, \href{http://arxiv.org/abs/1105.1333}{{\ttfamily arXiv:1105.1333
  [hep-th]}}.

\bibitem{Spradlin:2011wp}
M.~Spradlin and A.~Volovich, ``{Symbols of One-Loop Integrals From Mixed Tate
  Motives},'' \href{http://dx.doi.org/10.1007/JHEP11(2011)084}{{\em JHEP}
  {\bfseries 11} (2011) 084}, \href{http://arxiv.org/abs/1105.2024}{{\ttfamily
  arXiv:1105.2024 [hep-th]}}.

\bibitem{Li2021}
Z.~Li and et~al \href{http://arxiv.org/abs/to appear}{{\ttfamily to appear}}.

\bibitem{goncharov2005galois}
A.~B. Goncharov {\em et~al.}, ``Galois symmetries of fundamental groupoids and
  noncommutative geometry,'' {\em Duke Mathematical Journal} {\bfseries 128}
  no.~2, (2005) 209--284.

\end{thebibliography}\endgroup
\end{document}